\documentclass[aps,prl,twocolumn,superscriptaddress,showpacs,preprintnumbers]{revtex4-1}
\usepackage[colorlinks=true, pdfstartview=FitV, linkcolor=red, citecolor=blue, urlcolor=black, pdftitle={},pdfauthor={Tomoya Hayata},pdfsubject={}, pdfkeywords={}]{hyperref}
\usepackage{graphicx}
\usepackage{setspace,bm,times}
\usepackage{latexsym,amssymb,amsmath,mathrsfs}
\usepackage{color}
\graphicspath{{./figure/}}

\makeatletter
\providecommand \@ifxundefined [1]{%
 \@ifx{#1\undefined}
}%
\providecommand \@ifnum [1]{%
 \ifnum #1\expandafter \@firstoftwo
 \else \expandafter \@secondoftwo
 \fi
}%
\providecommand \@ifx [1]{%
 \ifx #1\expandafter \@firstoftwo
 \else \expandafter \@secondoftwo
 \fi
}%
\providecommand \bibnamefont  [1]{#1}%
\providecommand \bibfnamefont [1]{#1}%
\providecommand \citenamefont [1]{#1}%
\providecommand \href@noop [0]{\@secondoftwo}%
\providecommand \href [0]{\begingroup \@sanitize@url \@href}%
\providecommand \@href[1]{\@@startlink{#1}\@@href}%
\providecommand \@@href[1]{\endgroup#1\@@endlink}%
\providecommand \@sanitize@url [0]{\catcode `\\12\catcode `\$12\catcode
  `\&12\catcode `\#12\catcode `\^12\catcode `\_12\catcode `\%12\relax}%
\providecommand \@@startlink[1]{}%
\providecommand \@@endlink[0]{}%
\providecommand \url  [0]{\begingroup\@sanitize@url \@url }%
\providecommand \@url [1]{\endgroup\@href {#1}{\urlprefix }}%
\providecommand \urlprefix  [0]{URL }%
\providecommand \Eprint [0]{\href }%
\providecommand \doibase [0]{http://dx.doi.org/}%
\providecommand \selectlanguage [0]{\@gobble}%
\providecommand \bibinfo  [0]{\@secondoftwo}%
\providecommand \bibfield  [0]{\@secondoftwo}%
\providecommand \BibitemOpen [0]{}%
\providecommand \EOS [0]{\spacefactor3000\relax}%
\providecommand \BibitemShut  [1]{\csname bibitem#1\endcsname}%
\let\auto@bib@innerbib\@empty

\newcommand{\cL}{{\cal L}}

\newcommand{\vphi}{\varphi}
\newcommand{\vPhi}{\varPhi}

\newcommand{\be}{\begin{equation}}      
\newcommand{\ee}{\end{equation}}      
\newcommand{\bea}{\begin{eqnarray}}      
\newcommand{\eea}{\end{eqnarray}}

\begin{document}
\title{Diffusive Nambu--Goldstone modes in quantum time-crystals}

\author{Tomoya Hayata}
\affiliation{
Department of Physics, Chuo University, 1-13-27 Kasuga, Bunkyo, Tokyo, 112-8551, Japan 
}
\author{Yoshimasa Hidaka}
\affiliation{
Nishina Center, RIKEN, Wako, Saitama 351-0198, Japan
}
\affiliation{
iTHEMS Program, RIKEN, Wako, Saitama 351-0198, Japan
}
\preprint{RIKEN-QHP-380}
\date{\today}

\begin{abstract}
We study the Nambu--Goldstone (NG) modes associated with spontaneous breaking of the continuous time-translation symmetry. 
To discuss a quantum time-crystal with the spontaneously-broken continuous time-translation symmetry, we introduce the van der Pol type nonlinear-friction to open quantum systems.
By considering small fluctuations around a time-periodic mean-field solution, we show that a gapless collective mode necessarily appears; this is nothing but the NG mode associated with 
a time crystal.
We show that its dispersion relation becomes $\omega=-iC\bm p^2$. 
We also show that noncommutative breaking of the time-translation and U(1) symmetries results in mixing of the NG modes, and the (typically) propagating NG mode appears, whose dispersion relation becomes $\omega=(\pm C_1-iC_2)\bm p^2$.

\end{abstract}

\maketitle

\paragraph{Introduction.} 
The Nambu--Goldstone (NG) modes are universal gapless modes associated with spontaneous symmetry breaking (SSB)~\cite{Nambu:1961fr,Goldstone:1961eq,Goldstone:1962es}; they determine the low-energy or long-distance behavior of order-parameter fields.
For SSB in equilibrium states, the properties of the NG modes are well understood, and determined only by symmetry breaking patterns.
In particular,  for spontaneous breaking of internal symmetries, the number of the NG modes is given as~\cite{Watanabe:2011ec,Hidaka:2012ym,Watanabe:2012hr,Watanabe:2014fva} 
\bea
N_{\rm NG} &=& N_{\rm A}+N_{\rm B} , \\
N_{\rm A} &=& N_{\rm BS}-2N_{\rm B} , \\
N_{\rm B} &=& \frac{1}{2}\mathrm{rank}\, \rho^{ab} ,
\eea
where $N_{\rm NG}$, $N_{\rm A}$, and $N_{\rm B}$ are the number of the total, type-A, and type-B NG modes, respectively.
$N_{\rm BS}$ is the number of broken symmetries (broken generators).
$\rho_{ab}$ is the Watanabe-Brauner matrix defined by $\rho_{ab}:=\langle [i\hat{Q}_a,\hat{Q}_b]\rangle/V$ with the volume of the system $V$~\cite{Watanabe:2011ec}.
$\hat{Q}^a$ are broken charge operators; the total number of them is equal to $N_{\rm BS}$.
$\langle\cdots\rangle$ means expectation values, and hat denotes quantum operators.
In recent studies~\cite{Leutwyler:1993gf,Schafer:2001bq,Watanabe:2011ec,Hidaka:2012ym,Watanabe:2012hr,Watanabe:2014fva}, it was revealed that noncommutativity of broken charges plays a key role in the dynamics of the NG modes, 
and the NG modes are classified into the commutative and noncommutative (type-A and type-B) ones. 
They have the different dispersion relations;
the dispersion relations of the type-A NG modes are typically linear, while those of the type-B NG modes are typically quadratic (See e.g, Ref.~\cite{Hayata:2014yga}).

In contrast, only a little is known for SSB in nonequilibrium states such as the steady state with driven-dissipative condensates~\cite{2007PhRvL..99n0402W,2016RPPh...79i6001S}. 
Recently in Ref.~\cite{Minami:2018oxl}, from the theoretical analysis of some concrete models, 
it was shown that the dispersion relations of the type-A and type-B NG modes may become
\bea
\omega_{\rm A}&=&-iC_{\rm A}\bm p^2 ,\\
\omega_{\rm B}&=&\left(\pm C_{{\rm B}1}-iC_{{\rm B}2}\right)\bm p^2 ,
\eea
where $\omega_{\rm A,B}$, $\bm p$, and $C_{{\rm A,B}i}$ are frequency, momentum, and positive constants, respectively.
While the type-A NG mode becomes diffusive, the type-B NG mode still propagates because of the nonzero real part.
The two types were still classified by noncommutativity of broken charges in Ref.~\cite{Minami:2018oxl}.

It has been intensively discussed generalizing the notion of SSB to cover the time-direction. 
Such a quantum many-body state spontaneously breaking the time-translation symmetry is referred to as ``quantum time crystal," since it is a temporal analogue of the conventional crystal spontaneously breaking the spatial-translation symmetry~\cite{Wilczek:2012jt}.
It was shown that quantum time crystals can be realized only in nonequilibirum~\cite{Watanabe:2014hea}.
Soon after Ref.~\cite{Watanabe:2014hea}, spontaneous breaking of a discrete time-translation symmetry in periodically-driven open quantum systems has been proposed from theories~\cite{2016PhRvL.116y0401K,2016PhRvL.117i0402E,2017PhRvL.118c0401Y} and confirmed by experiments~\cite{2017Natur.543..217Z,2017Natur.543..221C,2018PhRvL.120r0603R,2017arXiv170808443P}. 

Spontaneous breaking of the continuous time-translation symmetry in the nonequilibirum Bose-Einstein condensate has also been discussed in literatures~\cite{Syrwid:2017tcx,PhysRevLett.120.215301} (A similar oscillatory condensate protected by the axial anomaly was discussed in Ref.~\cite{Hayata:2013sea}).
However, only the coherent phase-precession, or equivalently,  the phase oscillation by the chemical potential in an excited state has been discussed, but it in a nonequilibrium steady state has not been discussed.
Furthermore, the phase oscillation breaks only one of the two superpositions of the continuous time-translation and U($1$) symmetries, 
and the dispersion relation of the NG mode (phonon) is the same as that of the non-oscillatory Bose-Einstein condensate;
the effect of the oscillation can be eliminated by the local U($1$) transformation, and only the non-oscillatory amplitude affects the properties of the phonon.
To discuss the NG modes unique to quantum time-crystals, we need the temporal oscillation of not the phase, but the amplitude.

In this Letter, we study the NG modes in quantum time-crystals with the spontaneously-broken continuous time-translation symmetry.
To realize such a quantum time crystal as a nonequilibirum steady state, 
we consider open quantum systems with the van der Pol type nonlinear friction.
As is known in classical mechanics under the name of ``limit cycle," the van der Pol type nonlinear friction enables an emergent time-periodic solution.
By studying small fluctuations around a time-crystalline state, we show the existence of the gapless (NG) modes associated with quantum time-crystals.
We analyze the dispersion relations of them by generalizing the effective Lagrangian method~\cite{Watanabe:2012hr,Watanabe:2014fva} to open systems.

\paragraph{Effective Lagrangian.}
Let us consider the Keldysh action of an open quantum system $S[\vPhi_R,\vPhi_A]=\int d^4x\cL$ with double fields $\vPhi_R$, and $\vPhi_A$ in the so called R/A basis~\cite{suppl} (see Ref.~\cite{2016RPPh...79i6001S} for a review).
For concreteness, we here assume that $\vPhi_{R,A}$ are complex scalar fields.
The expectation value of an operator $O$ is given by the path integral:
\bea
\langle O \rangle  = \int \mathcal{D}\vPhi_{R}\mathcal{D}\vPhi_{R}^*\mathcal{D}\vPhi_{A}\mathcal{D}\vPhi_{A}^*e^{iS}O.
\eea
We assume the action is invariant under some transformations, $\vPhi_{R}\to \vPhi_{R}+\epsilon\delta_{A}^{a}\vPhi_{R}$
and $\vPhi_{A}\to \vPhi_{A}+\epsilon\delta_{A}^{a}\vPhi_{A}$, where $a$ labels symmetries, and $\epsilon$ is an infinitesimal small parameter.
For example, the time-translation and $U(1)$ transformations are defined as
 $\delta_{A}^{T}\vPhi_{R}:= \partial_{t}\vPhi_{R}, \delta_{A}^{T}\vPhi_{A}:= \partial_{t}\vPhi_{A}$, 
 and $\delta_{A}^{Q}\vPhi_{R}:=i \vPhi_{R}$, $\delta_{A}^{Q}\vPhi_{A}:=i \vPhi_{A}$. 
 The Noether theorem tells us the existence of the Noether currents $j_{A}^{a\mu}$ satisfying the conservation laws $\partial_{\mu}j_{A}^{a\mu}=0$. Hereafter, the Einstein convention is assumed for the repeated indices.
In the operator formalism, $\hat{Q}^{a}_{A}=\int d^{3}x \hat{j}^{a0}_A$ are the generators of the symmetries, and satisfy $-i[\hat{Q}^{a}_{A},\hat{\vPhi}_{R,A}]=\delta_{A}^{a}\hat{\vPhi}_{R,A}$.
Then, one might be tempted to use  $\langle[i\hat{Q}_{A}^{a},\hat{j}_{A}^{b0}]\rangle=\langle\delta^{a}_{A}\hat{j}_{A}^{b0}\rangle$ for the classification of the NG modes. However, $\langle\delta^{a}_{A}j_{A}^{b0}\rangle$ always vanish due to the conservation of probability~\cite{2016RPPh...79i6001S}. Instead, we introduce the transformations such that 
$\delta_{R}^{a}\vPhi_{R}:=\delta_{A}^{a}\vPhi_{A}/4$, $\delta_{R}^{a}\vPhi_{A}:=\delta_{A}^{a}\vPhi_{R}$~\cite{Minami:2018oxl,suppl}, which are
 the exact symmetries in isolated systems, while not in open systems. 
Nevertheless, $\langle \delta_{R}^{a}\hat{j}_{A}^{0b}\rangle$ plays a key role in the low-energy effective theory of the NG modes.
To explain this, let us remind the conventional NG mode associated with spontaneous breaking of a U(1) symmetry such as a phonon in superfluids.
The broken U(1) phase is characterized by a complex-scalar condensate $\vPhi_0$, and ground states are infinitely degenerate under the transformation $\vPhi_0\rightarrow e^{i\chi}\vPhi_0$ with constant $\chi$.
By promoting $\chi$ to the dynamical variable, we can introduce the dynamical fields for describing the phonon. 
The systematic construction of the effective theory of the NG modes along this line on the basis of the coset space of symmetry groups is the so called effective Lagrangian method~\cite{Watanabe:2012hr,Watanabe:2014fva}.
We can derive the effective theory of the NG modes in open systems in the same way~\cite{Hongo:2018ant}.
Following the equilibrium case, the dynamical fields of the NG modes can be introduced via the symmetry transformations as 
\bea
\vPhi_{R}&& =\langle\vPhi_{R}\rangle+\langle\delta^{a}_{A}\vPhi_{R}\rangle\chi_{a}+h_{R}, \quad \notag\\
\vPhi_{A}&&=\langle\delta^{a}_{R}\vPhi_{A}\rangle\pi_{a}+h_{A}.
\label{eq:NG_U(1)}
\eea
Here, $\chi_{a}$ and $\pi_{a}$ represent the NG fields, which are the degrees of freedom of the low-energy effective theory.
$h_{R,A}$ represent the non-NG fields, which will be integrated out to obtain the effective Lagrangian.
We note that $\langle\vPhi_{A}\rangle=0$ always holds.
In this parameterization of the fields, the quadratic part of the Lagrangian reads 
\bea
\begin{split}
&\cL =
\\
& \rho^{ab}\pi_{b}\partial_{t}\chi_{b} + 
 g_{t}^{ab}\partial_{t}\pi_{a}\partial_{t}\chi_{b}-g_{s}^{ab}\nabla\pi_{a}\cdot\nabla\chi_{b} + \frac{i}{2} g_{f}^{ab}\pi_{a}\pi_{b} ,
\end{split}
\label{eq:EFT}
\eea
where the coefficients are explicitly evaluated in our model introduced below as
\bea
\rho^{ab} &=& -\langle\delta_{R}^{a}j_{A}^{b0}\rangle,  \label{eq:coefficients1}\\
g_{t}^{ab} &=& \langle\delta_{A}^{a}\vPhi^{*}_{R}\rangle\langle\delta_{A}^{b}\vPhi_{R}\rangle+
\langle\delta_{A}^{a}\vPhi_{R}\rangle\langle\delta_{A}^{b}\vPhi^{*}_{R}\rangle,\label{eq:coefficients2}\\
g_{s}^{ab} &=& g_{t}^{ab},\\
g_{f}^{ab} &=& 2Ag_{t}^{ab},
\label{eq:coefficients}
\eea
and $\rho^{ab}$ is a generalization of the Watanabe-Brauner matrix for isolated systems at zero temperature~\cite{Watanabe:2011ec} to open quantum systems.
As expected, the first-order time-derivative terms in the Lagrangian~\eqref{eq:EFT} are determined by $\langle\delta_{R}^{a}j_{A}^{b0}\rangle$.
From the Lagrangian~\eqref{eq:EFT}, the linearized equation of motion of the NG fields is given as $(\rho^{ab}\partial_{t}-(\partial_{t} g_{t}^{ab})\partial_{t}- g_{t}^{ab}\partial_{t}^{2}+ g_{s}^{ab} \nabla^{2})\chi_{b}=-ig_{f}^{ab}\pi_{b}$.
Unlike isolated systems, $\rho^{ab}$ can be different from $\langle[i\hat{Q}_{R}^{a},\hat{j}_{A}^{b0}]\rangle$ because $\hat{Q}_{R}^{a}$ is not conserved.
In general, $\rho^{ab}$ is not anti-symmetric; the symmetric components represent friction.
The usual propagating type-B NG modes still appear when the anti-symmetric components of $\rho^{ab}$ do not vanish~\cite{Minami:2018oxl}.
Below we apply the effective Lagrangian~\eqref{eq:EFT} to quantum time-crystals.
Equation~\eqref{eq:coefficients1} is universal, although it was evaluated at the tree-level~\cite{Minami:2018}.
In contrast,  Eqs.~\eqref{eq:coefficients2}-\eqref{eq:coefficients} will be modified by fluctuations beyond the tree-level.

\paragraph{Van der Pol oscillator.}
We consider a field-theoretical generalization of the van der Pol oscillator with kinetic and interaction terms.
The Keldysh action of the corresponding open-quantum system is given as $S=\int d^4x\;\cL$, with
\bea
\begin{split}
\cL=&\vphi_A \left(-\partial^2_t+\nabla^2+\gamma\left(1-\kappa\vphi_R^2\right)\partial_t-2\lambda\vphi_R^2\right)\vphi_R\;\;
 \\
&\quad+i A(\vphi_A)^2 -\frac{1}{2}\lambda \vphi_{A}^{3}\vphi_{R},
\end{split}
\label{eq:action_van}
\eea
where $\vphi_{R,A}$ are real scalar fields in the R/A basis, $\gamma$, $\kappa>0$ are coefficients of the linear and cubic frictions, $\lambda$ is the coupling strength of the four-point contact interaction, and $A$ is the coefficient of fluctuations, respectively. 
We consider the nonlinear friction $-\gamma(1 -\kappa\vphi_R^2)\partial_t\vphi_R$, 
where $\gamma$ takes the negative sign compared to the conventional friction.
Therefore, $-\gamma(1 -\kappa\vphi_R^2)\partial_t\vphi_R$ works as the driving force if $\vphi_R<\sqrt{1/\kappa}$, while it works as the dissipative friction if $\vphi_R>\sqrt{1/\kappa}$. 
This is the characteristic of the van der Pol oscillator; because of the negative linear friction, uniform states are unstable against the temporal oscillation.
Since it is hard to exactly solve the real-time evolution of quantum many-body systems, we analyze Eq.~\eqref{eq:action_van} within the mean-field approximation.
\begin{figure}[t]
\centering
 \includegraphics[width=.4\textwidth]{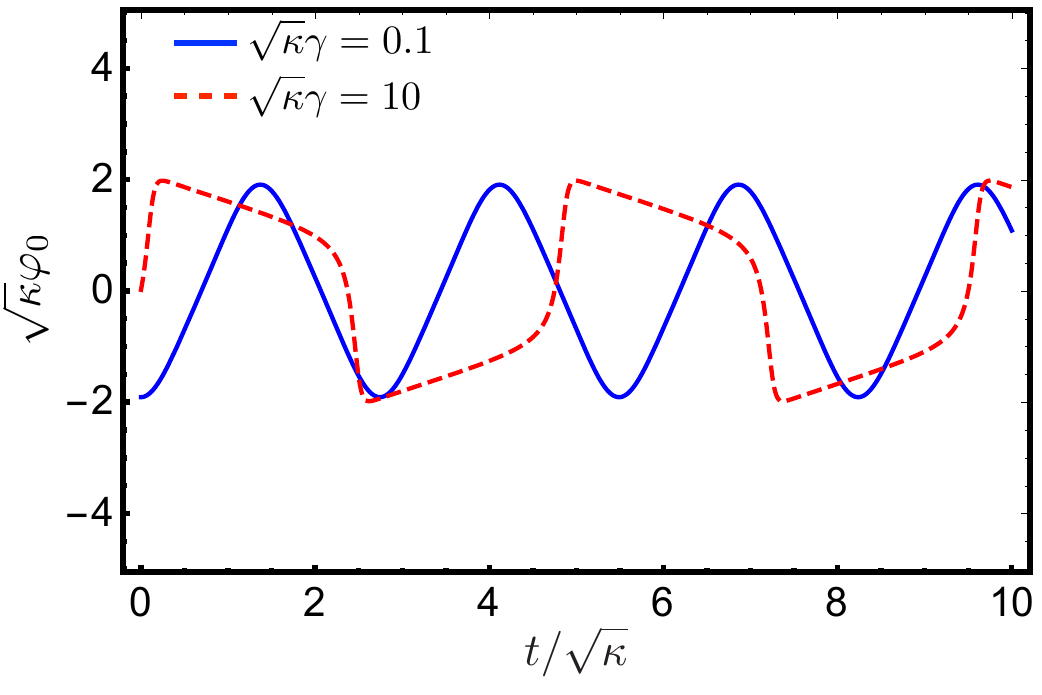}
 \caption{
Time-periodic solution of Eq.~\eqref{eq:eom_van}.
Blue solid and red dashed curves show $\vphi_0(t)$ at $\lambda=1.0$, and $\sqrt{\kappa}\gamma=0.1$, $10$.
}
\label{fig1}
\centering
 \includegraphics[width=.4\textwidth]{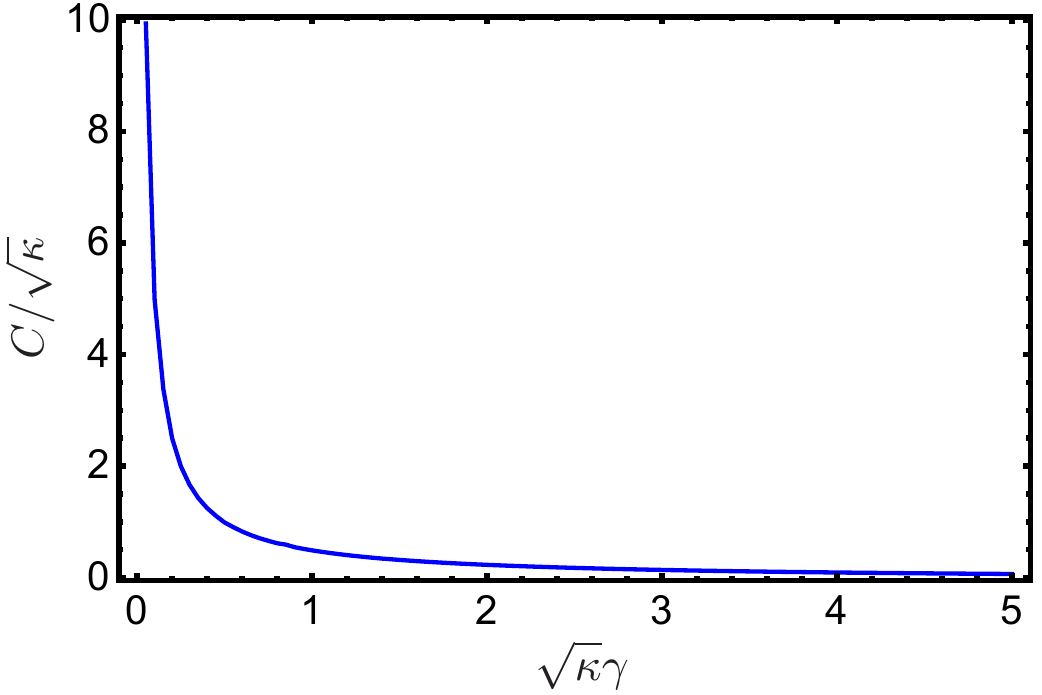}
 \caption{
Diffusive coefficient $C$ as a function of $\gamma$ at $\lambda=1.0$.
}
\label{fig2}
\end{figure}

We assume the spatially uniform solutions like the conventional mean-field theory.
Then, the classical saddles satisfy
\bea
-\partial^2_t\vphi_R+\gamma\left(1-\kappa\vphi_R^2\right)\partial_t\vphi_R-2\lambda\vphi_R^3=0 ,
\label{eq:eom_van}
\eea
and $\vphi_A(t)=0$. 
The classical value of $\vphi_R(t)$ exhibits a time-periodic solution $\vphi_0(t)$ in nonequilibrium steady states as shown in Fig.~\ref{fig1}, in addition to the unstable static solution $\vphi_R(t)=0$.
$\vphi_0(t)$ shows sinusoidal curves at $\sqrt{\kappa}\gamma\ll1$, while it  shows periodic spikes at $\sqrt{\kappa}\gamma\gtrsim1$ as shown in Fig.~\ref{fig1}.
In both cases, $\vphi_0(t)$ spontaneously breaks the time-translation symmetry.
Thus, we would realize a quantum time-crystal with spontaneous breaking of the continuous time-translation symmetry 
as the driven-dissipative condensate with the van der Pol type friction.
Below, we study fluctuations around the quantum time-crystal in detail, and the practical implementation of Eq.~\eqref{eq:eom_van} in a real system is left for a future study.

We consider a small fluctuation around $\vphi_0(t)$, and compute the dispersion relation of the NG mode.
$\vphi_0(t)$ has the degeneracy with respect to constant temporal shifts: $\vphi_0(t)\rightarrow\vphi_0(t+\chi)$.
Then, we promote $\chi$ to the dynamical variable. By substituting $\vphi_R(t,\bm x)=\vphi_0(t)+\dot{\vphi}_0(t)\chi(t,\bm x)$, and $\vphi_A(t,\bm x)=\dot{\vphi}_0(t)\pi(t,\bm x)$ into Eq.~\eqref{eq:action_van}~\footnote{Here, we expand the field with respect to $\chi$ and $\pi$ up to linear order because we are interested in the dispersion relation of NG modes.
We note that we do not need to introduce $h_{R,A}$ fields in the van del Pol oscillator. Since $\vphi_{R,A}$ are real scalar fields, the degrees of freedom is one. It is enough to parametrize the fields by $\chi$ and $\pi$. As the same reason, we do not need to introduce $h_{R,A}$ for the complex scalar model, where the fields are completely parametrized by $\chi^{T,Q}$ and $\pi^{T,Q}$.
}, we obtain the real-scalar-field version of the effective Lagrangian~\eqref{eq:EFT} with the coefficients
$g=\dot{\vphi}_{0}^{2}$, and $\rho=\dot{\vphi}_{0}^{2}\gamma(1-\kappa\vphi_{0}^{2})$, where the dot mean the time derivatives~\cite{suppl}.
The linearized equation of motion reads $(-g\partial_{t}^{2}+g\nabla^{2}+(\rho-\dot{g})\partial_{t} )\chi=:D_R^{-1}(t,\partial_t,\nabla)\chi=-2iA\dot{\vphi}_{0}^{2}\pi$.
The dispersion relation of the NG mode is obtained from it ($\pi=0$) as  
\bea
\omega=-iC\bm p^2 ,
\label{eq:typeA}
\eea
with
\bea
C^{-1}=\int_0^Tdt{\psi}_0^{\dag} \left(\dot{g} -\rho\right),
\eea
where $T$,  and $\psi_0$ are the period of $\vphi_0(t)$, and the zero eigen-function of the adjoint of $D^{-1}_{R}$: $(D^{-1}_R)^{\dag}{\psi}_0 =0$
normalized as $\int_0^Tdtg{\psi}_0^{\dag}=1$~\cite{suppl}.
We show $C$ as a function of $\gamma$ in Fig.~\ref{fig2}.
As of the characteristic of the NG mode in driven-dissipative condensates, the NG mode becomes diffusive.

\paragraph{Superfluid kink time-crystal.}
To cover more general quantum systems, we introduce the van der Pol type friction to a theory of a complex scalar field. 
We study an interplay between the spontaneously broken time-translation and U($1$) symmetries in mind of the type-B NG modes~\cite{Minami:2018oxl}.
We discuss a model of a superfluid with particle gain and loss, whose Keldysh action is given as $S=\int d^4x\;\cL$, with
\bea
\begin{split}
\cL&=\vPhi_A^* \Bigl(-\left(\partial_t+i\mu\right)^2+\nabla^2
\\
&\qquad+\gamma\left(1-\kappa|\vPhi_R|^2\right)\partial_t-m^2-\lambda|\partial_t\vPhi_R|^2\Bigr)\vPhi_R\;\;\;\;
 \\
&\qquad+\Bigl({\rm Hermite\;\; conjugates}\Bigr)+i A|\vPhi_A|^2 ,
\label{eq:action_U(1)}
\end{split}
\eea
where $\vPhi_{R,A}$ are complex scalar fields, $\mu$ is the chemical potential for a U(1) symmetry, and $\gamma$, $\kappa>0$ are the coefficients of the van der Pol type nonlinear friction, respectively. 
Here, the nonlinear friction $-\gamma(1 -\kappa|\vPhi_R|^2)\partial_t\vPhi_R$ means particle gain and loss;
if $|\vPhi_R|<\sqrt{1/\kappa}$, the system obtains particles from the reservoir, 
while if $|\vPhi_R|>\sqrt{1/\kappa}$, the system loses particles. 
We here consider the derivative interaction $\lambda\vPhi_A^*|\partial_t\vPhi_R|^2\vPhi_R$, instead of the conventional contact interaction $\lambda\vPhi_A^*|\vPhi_R|^2\vPhi_R$.
This is because the model with the latter interaction shows an unstable mode although the mean field solution shows even quantitatively similar behaviors with Eq.~\eqref{eq:action_U(1)}. 
The unstable mode implies that a spatiotemporal pattern is more favored than a spatially-uniform oscillatory pattern. 
However, since we are interested in a pure time-crystal, we introduce the derivative interaction to stabilize it.
Spatiotemporal patterns, that is, spacetime crystals, which might be widely favored than a pure time-crystal, and the associated NG modes will be studied in a future study.
We analyze Eq.~\eqref{eq:action_U(1)} with the spatially-uniform mean-field approximation.

\begin{figure}[t]
\centering
 \includegraphics[width=.4\textwidth]{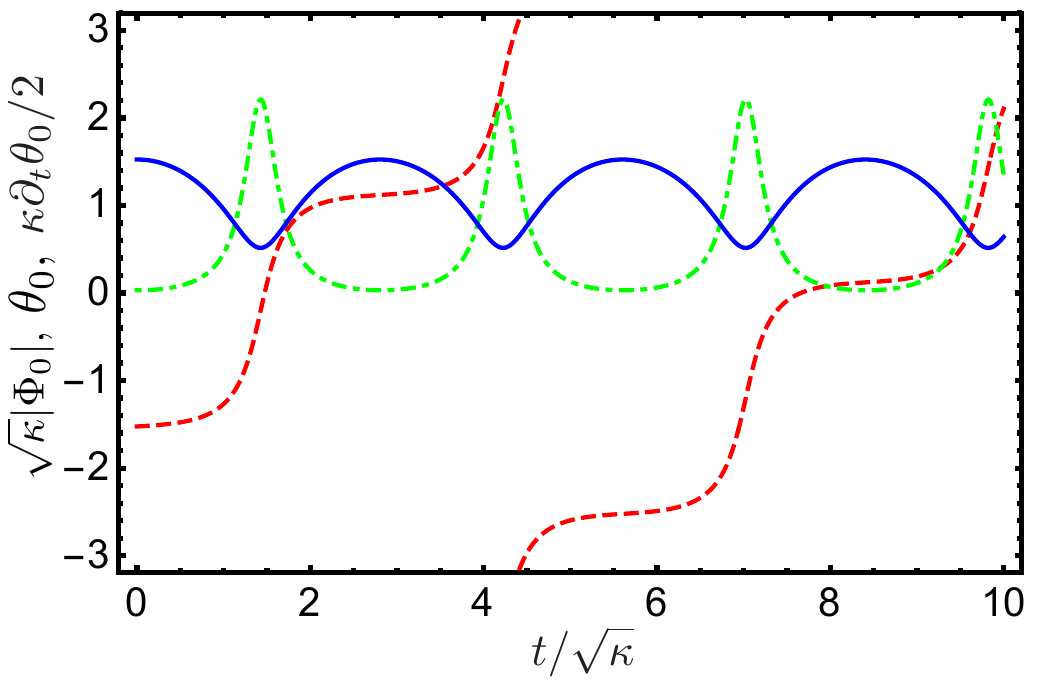}
 \caption{
Time-periodic solution of Eq.~\eqref{eq:eom_U(1)} at $\sqrt{\kappa}\gamma=0.1$, $\mu/\sqrt{\kappa}=0.5$, $\kappa m^2=1.0$, and $\lambda/\kappa=1.0$.
Blue solid, red dashed, and green dotdashed curves show $|\vPhi_0(t)|$, $\theta_0(t)$, and $\partial_t\theta_0(t)/2$, respectively.
}
\label{fig3}
\centering
 \includegraphics[width=.42\textwidth]{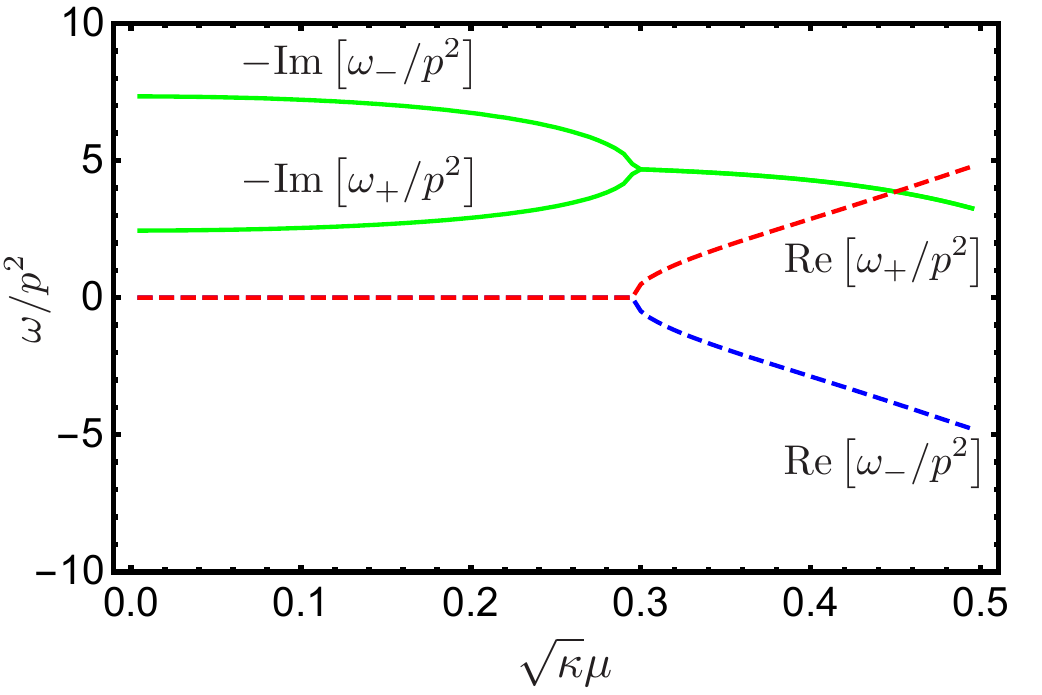}
 \caption{
Dispersion relation of the NG mode associated with the noncommutative broken time-translation and U(1) symmetries as a function of $\mu$ at $\sqrt{\kappa}\gamma=0.1$, $\kappa m^2=1.0$, and $\lambda/\kappa=1.0$.
}
\label{fig4}
\end{figure}

The classical saddles of Eq.~\eqref{eq:action_U(1)} satisfy
\bea
\begin{split}
-\left(\partial_t+i\mu\right)^2\vPhi_R&+\gamma\left(1-\kappa|\vPhi_R|^2\right)\partial_t\vPhi_R
\\
&-(m^2+\lambda|\partial_t\vPhi_R|^2)\vPhi_R=0 ,
\end{split}
\label{eq:eom_U(1)}
\eea
and $\vPhi_A=0$. 
The classical value of $\vPhi_R(t)$ exhibits a time-periodic solution $\vPhi_0(t)$ as shown in Fig.~\ref{fig3}. 
It spontaneously breaks both of the time-translation and U($1$) symmetries, and would be a temporal analogue of the twisted kink crystal~\cite{Basar:2008im}; 
the periodic time-lattice of kinks is formed as seen in Fig.~\ref{fig3}.
Therefore, we refer the state as ``kink time-crystal." 

We consider small fluctuations around $\vPhi_0(t)$, and compute the dispersion relation of the NG mode in the kink time-crystal; 
this can be done in the completely same way with the above real scalar field.
$\vPhi_0(t)$ has degeneracy with respect to constant temporal and U(1) shifts: $\vPhi_0(t)\rightarrow\vPhi_0(t+\chi_T)$, and $\vPhi_0(t)\rightarrow e^{i\chi_Q}\vPhi_0(t)$.
Then, by promoting $\chi_T$ and $\chi_Q$ to the dynamical variables via the symmetry transformations in Eq.~\eqref{eq:NG_U(1)}, we obtain the effective Lagrangian~\eqref{eq:EFT}, and linearized equation of motion of $\chi_{T}$ and $\chi_Q$. 
The linearized equation of motion necessarily contains temporal or spatial derivatives, so that $\chi_{T,Q}$ have gapless modes~\cite{suppl}, where two NG modes seem to appear.
However, they are not independent of each other, and form the one type-B NG mode.  
Since there are two variables, the dispersion relations are obtained from the determent of the $2\times2$ matrix:
\bea
{\rm det}
\begin{pmatrix}
\omega^2+ia\omega-\bm p^2 & -ic\omega \\
ib\omega & \omega^2+id\omega-\bm p^2
\end{pmatrix}=0.
\label{eq:det}
\eea
where the coefficients $a$, $b$, $c$, and $d$ need to be numerically evaluated~\cite{suppl}.
If $b$ or $c$ vanishes,  there are two diffusive NG modes ($\omega \sim -i\bm{p}^{2}$) like the previous case.
In contrast, if both of $b$ and $c$ are nonvanishing,  we have
\bea
\omega_{\pm}=\left(\pm C_1-iC_2\right)\bm p^2,
\label{eq:typeB}
\eea
where $C_{1}=(4bc-(a-d)^{2})^{1/2}/(2(ad+bc))$, and $C_{2}=(a+d)/(2(ad+bc))$.
This is the type-B NG mode associated with mixing of  the time-translation and U(1) symmetries.
We show $\omega_{\pm}/\bm p^2$ in Fig.~\ref{fig4}.
We see two diffusive NG modes below $\sqrt{\kappa}\mu=0.3$. 
However, they are in fact the overdamped one type-B NG mode; 
$C_1$ is pure-imaginary below $\sqrt{\kappa}\mu=0.3$, where Eq.~\eqref{eq:typeB} accidentally looks like two diffusive NG modes.
As $\mu$ increases, $C_1$ becomes real, and the propagating type-B NG mode appears as shown in Fig.~\ref{fig4}.
We note that the overdamping does not occur in isolated systems,
where  $C_{1}$ is always real for enough small $\bm p^{2}$~\cite{Hayata:2014yga}.
Such an overdamped type-B NG mode might be characteristic of driven-dissipative condensates.

\paragraph{Summary.}
We have studied the NG modes associated with spontaneous breaking of the continuous time-translation symmetry.
To realize a quantum time-crystal as a nonequilibirum steady state, we introduced the van der Pol type nonlinear-friction to open quantum systems.
We analyzed the dispersion relations of the collective modes around a time-crystalline state, and found that the gapless mode necessarily exists because of the temporal-degeneracy of the periodic states.
This is nothing but the NG mode associated with spontaneous breaking of the continuous time-translation symmetry.
We show that the NG mode from the broken time-translation symmetry behaves as $\omega=-iC\bm p^2$. 
We also show that noncommutative breaking of the time-translation and U(1) symmetries leads to the (typically) propagating NG mode; its dispersion relation becomes $\omega=(\pm C_1-iC_2)\bm p^2$.

We found the instability implying a spatially inhomogeneous order in the complex scalar model with the contact potential. 
It is interesting to discuss spacetime crystals and the NG modes associated with simultaneously broken time- and spatial-translation symmetries in a future study.
Such a study might be relevant for a phenomenological realization of the van der Pol type quantum time-crystal.

\begin{acknowledgements}
This work was supported by Japan Society of Promotion of Science (JSPS) Grant-in-Aid for Scientific Research 
Grants No. JP16J02240, 16K17716, 17H06462, and 18H01211.
Y. H. was  partially supported by RIKEN iTHES Project and iTHEMS Program.
\end{acknowledgements}

\bibliography{./ng}

\begin{thebibliography}{30}%
\makeatletter
\providecommand \@ifxundefined [1]{%
 \@ifx{#1\undefined}
}%
\providecommand \@ifnum [1]{%
 \ifnum #1\expandafter \@firstoftwo
 \else \expandafter \@secondoftwo
 \fi
}%
\providecommand \@ifx [1]{%
 \ifx #1\expandafter \@firstoftwo
 \else \expandafter \@secondoftwo
 \fi
}%
\providecommand \natexlab [1]{#1}%
\providecommand \enquote  [1]{``#1''}%
\providecommand \bibnamefont  [1]{#1}%
\providecommand \bibfnamefont [1]{#1}%
\providecommand \citenamefont [1]{#1}%
\providecommand \href@noop [0]{\@secondoftwo}%
\providecommand \href [0]{\begingroup \@sanitize@url \@href}%
\providecommand \@href[1]{\@@startlink{#1}\@@href}%
\providecommand \@@href[1]{\endgroup#1\@@endlink}%
\providecommand \@sanitize@url [0]{\catcode `\\12\catcode `\$12\catcode
  `\&12\catcode `\#12\catcode `\^12\catcode `\_12\catcode `\%12\relax}%
\providecommand \@@startlink[1]{}%
\providecommand \@@endlink[0]{}%
\providecommand \url  [0]{\begingroup\@sanitize@url \@url }%
\providecommand \@url [1]{\endgroup\@href {#1}{\urlprefix }}%
\providecommand \urlprefix  [0]{URL }%
\providecommand \Eprint [0]{\href }%
\providecommand \doibase [0]{http://dx.doi.org/}%
\providecommand \selectlanguage [0]{\@gobble}%
\providecommand \bibinfo  [0]{\@secondoftwo}%
\providecommand \bibfield  [0]{\@secondoftwo}%
\providecommand \translation [1]{[#1]}%
\providecommand \BibitemOpen [0]{}%
\providecommand \bibitemStop [0]{}%
\providecommand \bibitemNoStop [0]{.\EOS\space}%
\providecommand \EOS [0]{\spacefactor3000\relax}%
\providecommand \BibitemShut  [1]{\csname bibitem#1\endcsname}%
\let\auto@bib@innerbib\@empty
\bibitem [{\citenamefont {Nambu}\ and\ \citenamefont
  {Jona-Lasinio}(1961)}]{Nambu:1961fr}%
  \BibitemOpen
  \bibfield  {author} {\bibinfo {author} {\bibfnamefont {Y.}~\bibnamefont
  {Nambu}}\ and\ \bibinfo {author} {\bibfnamefont {G.}~\bibnamefont
  {Jona-Lasinio}},\ }\href {\doibase 10.1103/PhysRev.124.246} {\bibfield
  {journal} {\bibinfo  {journal} {Phys. Rev.}\ }\textbf {\bibinfo {volume}
  {124}},\ \bibinfo {pages} {246} (\bibinfo {year} {1961})},\ \bibinfo {note}
  {[,141(1961)]}\BibitemShut {NoStop}%
\bibitem [{\citenamefont {Goldstone}(1961)}]{Goldstone:1961eq}%
  \BibitemOpen
  \bibfield  {author} {\bibinfo {author} {\bibfnamefont {J.}~\bibnamefont
  {Goldstone}},\ }\href {\doibase 10.1007/BF02812722} {\bibfield  {journal}
  {\bibinfo  {journal} {Nuovo Cim.}\ }\textbf {\bibinfo {volume} {19}},\
  \bibinfo {pages} {154} (\bibinfo {year} {1961})}\BibitemShut {NoStop}%
\bibitem [{\citenamefont {Goldstone}\ \emph {et~al.}(1962)\citenamefont
  {Goldstone}, \citenamefont {Salam},\ and\ \citenamefont
  {Weinberg}}]{Goldstone:1962es}%
  \BibitemOpen
  \bibfield  {author} {\bibinfo {author} {\bibfnamefont {J.}~\bibnamefont
  {Goldstone}}, \bibinfo {author} {\bibfnamefont {A.}~\bibnamefont {Salam}}, \
  and\ \bibinfo {author} {\bibfnamefont {S.}~\bibnamefont {Weinberg}},\ }\href
  {\doibase 10.1103/PhysRev.127.965} {\bibfield  {journal} {\bibinfo  {journal}
  {Phys. Rev.}\ }\textbf {\bibinfo {volume} {127}},\ \bibinfo {pages} {965}
  (\bibinfo {year} {1962})}\BibitemShut {NoStop}%
\bibitem [{\citenamefont {Watanabe}\ and\ \citenamefont
  {Brauner}(2011)}]{Watanabe:2011ec}%
  \BibitemOpen
  \bibfield  {author} {\bibinfo {author} {\bibfnamefont {H.}~\bibnamefont
  {Watanabe}}\ and\ \bibinfo {author} {\bibfnamefont {T.}~\bibnamefont
  {Brauner}},\ }\href {\doibase 10.1103/PhysRevD.84.125013} {\bibfield
  {journal} {\bibinfo  {journal} {Phys. Rev.}\ }\textbf {\bibinfo {volume}
  {D84}},\ \bibinfo {pages} {125013} (\bibinfo {year} {2011})},\ \Eprint
  {http://arxiv.org/abs/1109.6327} {arXiv:1109.6327 [hep-ph]} \BibitemShut
  {NoStop}%
\bibitem [{\citenamefont {Hidaka}(2013)}]{Hidaka:2012ym}%
  \BibitemOpen
  \bibfield  {author} {\bibinfo {author} {\bibfnamefont {Y.}~\bibnamefont
  {Hidaka}},\ }\href {\doibase 10.1103/PhysRevLett.110.091601} {\bibfield
  {journal} {\bibinfo  {journal} {Phys. Rev. Lett.}\ }\textbf {\bibinfo
  {volume} {110}},\ \bibinfo {pages} {091601} (\bibinfo {year} {2013})},\
  \Eprint {http://arxiv.org/abs/1203.1494} {arXiv:1203.1494 [hep-th]}
  \BibitemShut {NoStop}%
\bibitem [{\citenamefont {Watanabe}\ and\ \citenamefont
  {Murayama}(2012)}]{Watanabe:2012hr}%
  \BibitemOpen
  \bibfield  {author} {\bibinfo {author} {\bibfnamefont {H.}~\bibnamefont
  {Watanabe}}\ and\ \bibinfo {author} {\bibfnamefont {H.}~\bibnamefont
  {Murayama}},\ }\href {\doibase 10.1103/PhysRevLett.108.251602} {\bibfield
  {journal} {\bibinfo  {journal} {Phys. Rev. Lett.}\ }\textbf {\bibinfo
  {volume} {108}},\ \bibinfo {pages} {251602} (\bibinfo {year} {2012})},\
  \Eprint {http://arxiv.org/abs/1203.0609} {arXiv:1203.0609 [hep-th]}
  \BibitemShut {NoStop}%
\bibitem [{\citenamefont {Watanabe}\ and\ \citenamefont
  {Murayama}(2014)}]{Watanabe:2014fva}%
  \BibitemOpen
  \bibfield  {author} {\bibinfo {author} {\bibfnamefont {H.}~\bibnamefont
  {Watanabe}}\ and\ \bibinfo {author} {\bibfnamefont {H.}~\bibnamefont
  {Murayama}},\ }\href {\doibase 10.1103/PhysRevX.4.031057} {\bibfield
  {journal} {\bibinfo  {journal} {Phys. Rev.}\ }\textbf {\bibinfo {volume}
  {X4}},\ \bibinfo {pages} {031057} (\bibinfo {year} {2014})},\ \Eprint
  {http://arxiv.org/abs/1402.7066} {arXiv:1402.7066 [hep-th]} \BibitemShut
  {NoStop}%
\bibitem [{\citenamefont {Leutwyler}(1994)}]{Leutwyler:1993gf}%
  \BibitemOpen
  \bibfield  {author} {\bibinfo {author} {\bibfnamefont {H.}~\bibnamefont
  {Leutwyler}},\ }\href {\doibase 10.1103/PhysRevD.49.3033} {\bibfield
  {journal} {\bibinfo  {journal} {Phys. Rev.}\ }\textbf {\bibinfo {volume}
  {D49}},\ \bibinfo {pages} {3033} (\bibinfo {year} {1994})},\ \Eprint
  {http://arxiv.org/abs/hep-ph/9311264} {arXiv:hep-ph/9311264 [hep-ph]}
  \BibitemShut {NoStop}%
\bibitem [{\citenamefont {Sch{\"a}fer}\ \emph {et~al.}(2001)\citenamefont
  {Sch{\"a}fer}, \citenamefont {Son}, \citenamefont {Stephanov}, \citenamefont
  {Toublan},\ and\ \citenamefont {Verbaarschot}}]{Schafer:2001bq}%
  \BibitemOpen
  \bibfield  {author} {\bibinfo {author} {\bibfnamefont {T.}~\bibnamefont
  {Sch{\"a}fer}}, \bibinfo {author} {\bibfnamefont {D.~T.}\ \bibnamefont
  {Son}}, \bibinfo {author} {\bibfnamefont {M.~A.}\ \bibnamefont {Stephanov}},
  \bibinfo {author} {\bibfnamefont {D.}~\bibnamefont {Toublan}}, \ and\
  \bibinfo {author} {\bibfnamefont {J.~J.~M.}\ \bibnamefont {Verbaarschot}},\
  }\href {\doibase 10.1016/S0370-2693(01)01265-5} {\bibfield  {journal}
  {\bibinfo  {journal} {Phys. Lett.}\ }\textbf {\bibinfo {volume} {B522}},\
  \bibinfo {pages} {67} (\bibinfo {year} {2001})},\ \Eprint
  {http://arxiv.org/abs/hep-ph/0108210} {arXiv:hep-ph/0108210 [hep-ph]}
  \BibitemShut {NoStop}%
\bibitem [{\citenamefont {Hayata}\ and\ \citenamefont
  {Hidaka}(2015)}]{Hayata:2014yga}%
  \BibitemOpen
  \bibfield  {author} {\bibinfo {author} {\bibfnamefont {T.}~\bibnamefont
  {Hayata}}\ and\ \bibinfo {author} {\bibfnamefont {Y.}~\bibnamefont
  {Hidaka}},\ }\href {\doibase 10.1103/PhysRevD.91.056006} {\bibfield
  {journal} {\bibinfo  {journal} {Phys. Rev.}\ }\textbf {\bibinfo {volume}
  {D91}},\ \bibinfo {pages} {056006} (\bibinfo {year} {2015})},\ \Eprint
  {http://arxiv.org/abs/1406.6271} {arXiv:1406.6271 [hep-th]} \BibitemShut
  {NoStop}%
\bibitem [{\citenamefont {{Wouters}}\ and\ \citenamefont
  {{Carusotto}}(2007)}]{2007PhRvL..99n0402W}%
  \BibitemOpen
  \bibfield  {author} {\bibinfo {author} {\bibfnamefont {M.}~\bibnamefont
  {{Wouters}}}\ and\ \bibinfo {author} {\bibfnamefont {I.}~\bibnamefont
  {{Carusotto}}},\ }\href {\doibase 10.1103/PhysRevLett.99.140402} {\bibfield
  {journal} {\bibinfo  {journal} {Phys. Rev. Lett.}\ }\textbf {\bibinfo
  {volume} {99}},\ \bibinfo {eid} {140402} (\bibinfo {year} {2007})},\ \Eprint
  {http://arxiv.org/abs/0702431} {arXiv:0702431 [cond-mat]} \BibitemShut
  {NoStop}%
\bibitem [{\citenamefont {{Sieberer}}\ \emph {et~al.}(2016)\citenamefont
  {{Sieberer}}, \citenamefont {{Buchhold}},\ and\ \citenamefont
  {{Diehl}}}]{2016RPPh...79i6001S}%
  \BibitemOpen
  \bibfield  {author} {\bibinfo {author} {\bibfnamefont {L.~M.}\ \bibnamefont
  {{Sieberer}}}, \bibinfo {author} {\bibfnamefont {M.}~\bibnamefont
  {{Buchhold}}}, \ and\ \bibinfo {author} {\bibfnamefont {S.}~\bibnamefont
  {{Diehl}}},\ }\href {\doibase 10.1088/0034-4885/79/9/096001} {\bibfield
  {journal} {\bibinfo  {journal} {Reports on Progress in Physics}\ }\textbf
  {\bibinfo {volume} {79}},\ \bibinfo {eid} {096001} (\bibinfo {year}
  {2016})},\ \Eprint {http://arxiv.org/abs/1512.00637} {arXiv:1512.00637
  [cond-mat.quant-gas]} \BibitemShut {NoStop}%
\bibitem [{\citenamefont {Minami}\ and\ \citenamefont
  {Hidaka}(2018)}]{Minami:2018oxl}%
  \BibitemOpen
  \bibfield  {author} {\bibinfo {author} {\bibfnamefont {Y.}~\bibnamefont
  {Minami}}\ and\ \bibinfo {author} {\bibfnamefont {Y.}~\bibnamefont
  {Hidaka}},\ }\href {\doibase 10.1103/PhysRevE.97.012130} {\bibfield
  {journal} {\bibinfo  {journal} {Phys. Rev.}\ }\textbf {\bibinfo {volume}
  {E97}},\ \bibinfo {pages} {012130} (\bibinfo {year} {2018})},\ \Eprint
  {http://arxiv.org/abs/1509.05042} {arXiv:1509.05042 [cond-mat.stat-mech]}
  \BibitemShut {NoStop}%
\bibitem [{\citenamefont {Wilczek}(2012)}]{Wilczek:2012jt}%
  \BibitemOpen
  \bibfield  {author} {\bibinfo {author} {\bibfnamefont {F.}~\bibnamefont
  {Wilczek}},\ }\href {\doibase 10.1103/PhysRevLett.109.160401} {\bibfield
  {journal} {\bibinfo  {journal} {Phys. Rev. Lett.}\ }\textbf {\bibinfo
  {volume} {109}},\ \bibinfo {pages} {160401} (\bibinfo {year} {2012})},\
  \Eprint {http://arxiv.org/abs/1202.2539} {arXiv:1202.2539 [quant-ph]}
  \BibitemShut {NoStop}%
\bibitem [{\citenamefont {Watanabe}\ and\ \citenamefont
  {Oshikawa}(2015)}]{Watanabe:2014hea}%
  \BibitemOpen
  \bibfield  {author} {\bibinfo {author} {\bibfnamefont {H.}~\bibnamefont
  {Watanabe}}\ and\ \bibinfo {author} {\bibfnamefont {M.}~\bibnamefont
  {Oshikawa}},\ }\href {\doibase 10.1103/PhysRevLett.114.251603} {\bibfield
  {journal} {\bibinfo  {journal} {Phys. Rev. Lett.}\ }\textbf {\bibinfo
  {volume} {114}},\ \bibinfo {pages} {251603} (\bibinfo {year} {2015})},\
  \Eprint {http://arxiv.org/abs/1410.2143} {arXiv:1410.2143
  [cond-mat.stat-mech]} \BibitemShut {NoStop}%
\bibitem [{\citenamefont {{Khemani}}\ \emph {et~al.}(2016)\citenamefont
  {{Khemani}}, \citenamefont {{Lazarides}}, \citenamefont {{Moessner}},\ and\
  \citenamefont {{Sondhi}}}]{2016PhRvL.116y0401K}%
  \BibitemOpen
  \bibfield  {author} {\bibinfo {author} {\bibfnamefont {V.}~\bibnamefont
  {{Khemani}}}, \bibinfo {author} {\bibfnamefont {A.}~\bibnamefont
  {{Lazarides}}}, \bibinfo {author} {\bibfnamefont {R.}~\bibnamefont
  {{Moessner}}}, \ and\ \bibinfo {author} {\bibfnamefont {S.~L.}\ \bibnamefont
  {{Sondhi}}},\ }\href {\doibase 10.1103/PhysRevLett.116.250401} {\bibfield
  {journal} {\bibinfo  {journal} {Phys. Rev. Lett.}\ }\textbf {\bibinfo
  {volume} {116}},\ \bibinfo {eid} {250401} (\bibinfo {year} {2016})},\ \Eprint
  {http://arxiv.org/abs/1508.03344} {arXiv:1508.03344 [cond-mat.dis-nn]}
  \BibitemShut {NoStop}%
\bibitem [{\citenamefont {{Else}}\ \emph {et~al.}(2016)\citenamefont {{Else}},
  \citenamefont {{Bauer}},\ and\ \citenamefont
  {{Nayak}}}]{2016PhRvL.117i0402E}%
  \BibitemOpen
  \bibfield  {author} {\bibinfo {author} {\bibfnamefont {D.~V.}\ \bibnamefont
  {{Else}}}, \bibinfo {author} {\bibfnamefont {B.}~\bibnamefont {{Bauer}}}, \
  and\ \bibinfo {author} {\bibfnamefont {C.}~\bibnamefont {{Nayak}}},\ }\href
  {\doibase 10.1103/PhysRevLett.117.090402} {\bibfield  {journal} {\bibinfo
  {journal} {Phys. Rev. Lett.}\ }\textbf {\bibinfo {volume} {117}},\ \bibinfo
  {eid} {090402} (\bibinfo {year} {2016})},\ \Eprint
  {http://arxiv.org/abs/1603.08001} {arXiv:1603.08001 [cond-mat.dis-nn]}
  \BibitemShut {NoStop}%
\bibitem [{\citenamefont {{Yao}}\ \emph {et~al.}(2017)\citenamefont {{Yao}},
  \citenamefont {{Potter}}, \citenamefont {{Potirniche}},\ and\ \citenamefont
  {{Vishwanath}}}]{2017PhRvL.118c0401Y}%
  \BibitemOpen
  \bibfield  {author} {\bibinfo {author} {\bibfnamefont {N.~Y.}\ \bibnamefont
  {{Yao}}}, \bibinfo {author} {\bibfnamefont {A.~C.}\ \bibnamefont {{Potter}}},
  \bibinfo {author} {\bibfnamefont {I.-D.}\ \bibnamefont {{Potirniche}}}, \
  and\ \bibinfo {author} {\bibfnamefont {A.}~\bibnamefont {{Vishwanath}}},\
  }\href {\doibase 10.1103/PhysRevLett.118.030401} {\bibfield  {journal}
  {\bibinfo  {journal} {Phys. Rev. Lett.}\ }\textbf {\bibinfo {volume} {118}},\
  \bibinfo {eid} {030401} (\bibinfo {year} {2017})},\ \Eprint
  {http://arxiv.org/abs/1608.02589} {arXiv:1608.02589 [cond-mat.dis-nn]}
  \BibitemShut {NoStop}%
\bibitem [{\citenamefont {{Zhang}}\ \emph {et~al.}(2017)\citenamefont
  {{Zhang}}, \citenamefont {{Hess}}, \citenamefont {{Kyprianidis}},
  \citenamefont {{Becker}}, \citenamefont {{Lee}}, \citenamefont {{Smith}},
  \citenamefont {{Pagano}}, \citenamefont {{Potirniche}}, \citenamefont
  {{Potter}}, \citenamefont {{Vishwanath}}, \citenamefont {{Yao}},\ and\
  \citenamefont {{Monroe}}}]{2017Natur.543..217Z}%
  \BibitemOpen
  \bibfield  {author} {\bibinfo {author} {\bibfnamefont {J.}~\bibnamefont
  {{Zhang}}}, \bibinfo {author} {\bibfnamefont {P.~W.}\ \bibnamefont {{Hess}}},
  \bibinfo {author} {\bibfnamefont {A.}~\bibnamefont {{Kyprianidis}}}, \bibinfo
  {author} {\bibfnamefont {P.}~\bibnamefont {{Becker}}}, \bibinfo {author}
  {\bibfnamefont {A.}~\bibnamefont {{Lee}}}, \bibinfo {author} {\bibfnamefont
  {J.}~\bibnamefont {{Smith}}}, \bibinfo {author} {\bibfnamefont
  {G.}~\bibnamefont {{Pagano}}}, \bibinfo {author} {\bibfnamefont {I.-D.}\
  \bibnamefont {{Potirniche}}}, \bibinfo {author} {\bibfnamefont {A.~C.}\
  \bibnamefont {{Potter}}}, \bibinfo {author} {\bibfnamefont {A.}~\bibnamefont
  {{Vishwanath}}}, \bibinfo {author} {\bibfnamefont {N.~Y.}\ \bibnamefont
  {{Yao}}}, \ and\ \bibinfo {author} {\bibfnamefont {C.}~\bibnamefont
  {{Monroe}}},\ }\href {\doibase 10.1038/nature21413} {\bibfield  {journal}
  {\bibinfo  {journal} {\nat}\ }\textbf {\bibinfo {volume} {543}},\ \bibinfo
  {pages} {217} (\bibinfo {year} {2017})},\ \Eprint
  {http://arxiv.org/abs/1609.08684} {arXiv:1609.08684 [quant-ph]} \BibitemShut
  {NoStop}%
\bibitem [{\citenamefont {{Choi}}\ \emph {et~al.}(2017)\citenamefont {{Choi}},
  \citenamefont {{Choi}}, \citenamefont {{Landig}}, \citenamefont {{Kucsko}},
  \citenamefont {{Zhou}}, \citenamefont {{Isoya}}, \citenamefont {{Jelezko}},
  \citenamefont {{Onoda}}, \citenamefont {{Sumiya}}, \citenamefont {{Khemani}},
  \citenamefont {{von Keyserlingk}}, \citenamefont {{Yao}}, \citenamefont
  {{Demler}},\ and\ \citenamefont {{Lukin}}}]{2017Natur.543..221C}%
  \BibitemOpen
  \bibfield  {author} {\bibinfo {author} {\bibfnamefont {S.}~\bibnamefont
  {{Choi}}}, \bibinfo {author} {\bibfnamefont {J.}~\bibnamefont {{Choi}}},
  \bibinfo {author} {\bibfnamefont {R.}~\bibnamefont {{Landig}}}, \bibinfo
  {author} {\bibfnamefont {G.}~\bibnamefont {{Kucsko}}}, \bibinfo {author}
  {\bibfnamefont {H.}~\bibnamefont {{Zhou}}}, \bibinfo {author} {\bibfnamefont
  {J.}~\bibnamefont {{Isoya}}}, \bibinfo {author} {\bibfnamefont
  {F.}~\bibnamefont {{Jelezko}}}, \bibinfo {author} {\bibfnamefont
  {S.}~\bibnamefont {{Onoda}}}, \bibinfo {author} {\bibfnamefont
  {H.}~\bibnamefont {{Sumiya}}}, \bibinfo {author} {\bibfnamefont
  {V.}~\bibnamefont {{Khemani}}}, \bibinfo {author} {\bibfnamefont
  {C.}~\bibnamefont {{von Keyserlingk}}}, \bibinfo {author} {\bibfnamefont
  {N.~Y.}\ \bibnamefont {{Yao}}}, \bibinfo {author} {\bibfnamefont
  {E.}~\bibnamefont {{Demler}}}, \ and\ \bibinfo {author} {\bibfnamefont
  {M.~D.}\ \bibnamefont {{Lukin}}},\ }\href {\doibase 10.1038/nature21426}
  {\bibfield  {journal} {\bibinfo  {journal} {\nat}\ }\textbf {\bibinfo
  {volume} {543}},\ \bibinfo {pages} {221} (\bibinfo {year} {2017})},\ \Eprint
  {http://arxiv.org/abs/1610.08057} {arXiv:1610.08057 [quant-ph]} \BibitemShut
  {NoStop}%
\bibitem [{\citenamefont {{Rovny}}\ \emph {et~al.}(2018)\citenamefont
  {{Rovny}}, \citenamefont {{Blum}},\ and\ \citenamefont
  {{Barrett}}}]{2018PhRvL.120r0603R}%
  \BibitemOpen
  \bibfield  {author} {\bibinfo {author} {\bibfnamefont {J.}~\bibnamefont
  {{Rovny}}}, \bibinfo {author} {\bibfnamefont {R.~L.}\ \bibnamefont {{Blum}}},
  \ and\ \bibinfo {author} {\bibfnamefont {S.~E.}\ \bibnamefont {{Barrett}}},\
  }\href {\doibase 10.1103/PhysRevLett.120.180603} {\bibfield  {journal}
  {\bibinfo  {journal} {Phys. Rev. Lett.}\ }\textbf {\bibinfo {volume} {120}},\
  \bibinfo {eid} {180603} (\bibinfo {year} {2018})},\ \Eprint
  {http://arxiv.org/abs/1802.00126} {arXiv:1802.00126 [quant-ph]} \BibitemShut
  {NoStop}%
\bibitem [{\citenamefont {Pal}\ \emph {et~al.}(2018)\citenamefont {Pal},
  \citenamefont {Nishad}, \citenamefont {Mahesh},\ and\ \citenamefont
  {Sreejith}}]{2017arXiv170808443P}%
  \BibitemOpen
  \bibfield  {author} {\bibinfo {author} {\bibfnamefont {S.}~\bibnamefont
  {Pal}}, \bibinfo {author} {\bibfnamefont {N.}~\bibnamefont {Nishad}},
  \bibinfo {author} {\bibfnamefont {T.~S.}\ \bibnamefont {Mahesh}}, \ and\
  \bibinfo {author} {\bibfnamefont {G.~J.}\ \bibnamefont {Sreejith}},\ }\href
  {\doibase 10.1103/PhysRevLett.120.180602} {\bibfield  {journal} {\bibinfo
  {journal} {Phys. Rev. Lett.}\ }\textbf {\bibinfo {volume} {120}},\ \bibinfo
  {pages} {180602} (\bibinfo {year} {2018})},\ \Eprint
  {http://arxiv.org/abs/1708.08443} {arXiv:1708.08443 [cond-mat.stat-mech]}
  \BibitemShut {NoStop}%
\bibitem [{\citenamefont {Syrwid}\ \emph {et~al.}(2017)\citenamefont {Syrwid},
  \citenamefont {Zakrzewski},\ and\ \citenamefont {Sacha}}]{Syrwid:2017tcx}%
  \BibitemOpen
  \bibfield  {author} {\bibinfo {author} {\bibfnamefont {A.}~\bibnamefont
  {Syrwid}}, \bibinfo {author} {\bibfnamefont {J.}~\bibnamefont {Zakrzewski}},
  \ and\ \bibinfo {author} {\bibfnamefont {K.}~\bibnamefont {Sacha}},\ }\href
  {\doibase 10.1103/PhysRevLett.119.250602} {\bibfield  {journal} {\bibinfo
  {journal} {Phys. Rev. Lett.}\ }\textbf {\bibinfo {volume} {119}},\ \bibinfo
  {pages} {250602} (\bibinfo {year} {2017})},\ \Eprint
  {http://arxiv.org/abs/1702.05006} {arXiv:1702.05006 [quant-ph]} \BibitemShut
  {NoStop}%
\bibitem [{\citenamefont {Autti}\ \emph {et~al.}(2018)\citenamefont {Autti},
  \citenamefont {Eltsov},\ and\ \citenamefont
  {Volovik}}]{PhysRevLett.120.215301}%
  \BibitemOpen
  \bibfield  {author} {\bibinfo {author} {\bibfnamefont {S.}~\bibnamefont
  {Autti}}, \bibinfo {author} {\bibfnamefont {V.~B.}\ \bibnamefont {Eltsov}}, \
  and\ \bibinfo {author} {\bibfnamefont {G.~E.}\ \bibnamefont {Volovik}},\
  }\href {\doibase 10.1103/PhysRevLett.120.215301} {\bibfield  {journal}
  {\bibinfo  {journal} {Phys. Rev. Lett.}\ }\textbf {\bibinfo {volume} {120}},\
  \bibinfo {pages} {215301} (\bibinfo {year} {2018})},\ \Eprint
  {http://arxiv.org/abs/1712.06877} {arXiv:1712.06877 [cond-mat.other]}
  \BibitemShut {NoStop}%
\bibitem [{\citenamefont {Hayata}\ \emph {et~al.}(2014)\citenamefont {Hayata},
  \citenamefont {Hidaka},\ and\ \citenamefont {Yamamoto}}]{Hayata:2013sea}%
  \BibitemOpen
  \bibfield  {author} {\bibinfo {author} {\bibfnamefont {T.}~\bibnamefont
  {Hayata}}, \bibinfo {author} {\bibfnamefont {Y.}~\bibnamefont {Hidaka}}, \
  and\ \bibinfo {author} {\bibfnamefont {A.}~\bibnamefont {Yamamoto}},\ }\href
  {\doibase 10.1103/PhysRevD.89.085011} {\bibfield  {journal} {\bibinfo
  {journal} {Phys. Rev.}\ }\textbf {\bibinfo {volume} {D89}},\ \bibinfo {pages}
  {085011} (\bibinfo {year} {2014})},\ \Eprint {http://arxiv.org/abs/1309.0012}
  {arXiv:1309.0012 [hep-ph]} \BibitemShut {NoStop}%
\bibitem [{sup()}]{suppl}%
  \BibitemOpen
  \href@noop {} {\ }\bibinfo {note} {See Supplemental Material at [URL will be
  inserted by publisher] for computational details}\BibitemShut {NoStop}%
\bibitem [{\citenamefont {Hongo}\ \emph {et~al.}()\citenamefont {Hongo},
  \citenamefont {Kim}, \citenamefont {Noumi},\ and\ \citenamefont
  {Ota}}]{Hongo:2018ant}%
  \BibitemOpen
  \bibfield  {author} {\bibinfo {author} {\bibfnamefont {M.}~\bibnamefont
  {Hongo}}, \bibinfo {author} {\bibfnamefont {S.}~\bibnamefont {Kim}}, \bibinfo
  {author} {\bibfnamefont {T.}~\bibnamefont {Noumi}}, \ and\ \bibinfo {author}
  {\bibfnamefont {A.}~\bibnamefont {Ota}},\ }\href@noop {} {\ }\Eprint
  {http://arxiv.org/abs/1805.06240} {arXiv:1805.06240 [hep-th]} \BibitemShut
  {NoStop}%
\bibitem [{\citenamefont {Minami}\ and\ \citenamefont
  {Hidaka}()}]{Minami:2018}%
  \BibitemOpen
  \bibfield  {author} {\bibinfo {author} {\bibfnamefont {Y.}~\bibnamefont
  {Minami}}\ and\ \bibinfo {author} {\bibfnamefont {Y.}~\bibnamefont
  {Hidaka}},\ }\href@noop {} {\bibinfo  {journal} {in preparation}\
  }\BibitemShut {NoStop}%
\bibitem [{Note1()}]{Note1}%
  \BibitemOpen
\bibfield  {journal} {  }\bibinfo {note} {Here, we expand the field with
  respect to $\chi $ and $\pi $ up to linear order because we are interested in
  the dispersion relation of NG modes. We note that we do not need to introduce
  $h_{R,A}$ fields in the van del Pol oscillator. Since $\varphi _{R,A}$ are
  real scalar fields, the degrees of freedom is one. It is enough to
  parametrize the fields by $\chi $ and $\pi $. As the same reason, we do not
  need to introduce $h_{R,A}$ for the complex scalar model, where the fields
  are completely parametrized by $\chi ^{T,Q}$ and $\pi ^{T,Q}$.}\BibitemShut
  {Stop}%
\bibitem [{\citenamefont {Basar}\ and\ \citenamefont
  {Dunne}(2008)}]{Basar:2008im}%
  \BibitemOpen
  \bibfield  {author} {\bibinfo {author} {\bibfnamefont {G.}~\bibnamefont
  {Basar}}\ and\ \bibinfo {author} {\bibfnamefont {G.~V.}\ \bibnamefont
  {Dunne}},\ }\href {\doibase 10.1103/PhysRevLett.100.200404} {\bibfield
  {journal} {\bibinfo  {journal} {Phys. Rev. Lett.}\ }\textbf {\bibinfo
  {volume} {100}},\ \bibinfo {pages} {200404} (\bibinfo {year} {2008})},\
  \Eprint {http://arxiv.org/abs/0803.1501} {arXiv:0803.1501 [hep-th]}
  \BibitemShut {NoStop}%
\end{thebibliography}%


\begin{thebibliography}{1}%
\makeatletter
\providecommand \@ifxundefined [1]{%
 \@ifx{#1\undefined}
}%
\providecommand \@ifnum [1]{%
 \ifnum #1\expandafter \@firstoftwo
 \else \expandafter \@secondoftwo
 \fi
}%
\providecommand \@ifx [1]{%
 \ifx #1\expandafter \@firstoftwo
 \else \expandafter \@secondoftwo
 \fi
}%
\providecommand \natexlab [1]{#1}%
\providecommand \enquote  [1]{``#1''}%
\providecommand \bibnamefont  [1]{#1}%
\providecommand \bibfnamefont [1]{#1}%
\providecommand \citenamefont [1]{#1}%
\providecommand \href@noop [0]{\@secondoftwo}%
\providecommand \href [0]{\begingroup \@sanitize@url \@href}%
\providecommand \@href[1]{\@@startlink{#1}\@@href}%
\providecommand \@@href[1]{\endgroup#1\@@endlink}%
\providecommand \@sanitize@url [0]{\catcode `\\12\catcode `\$12\catcode
  `\&12\catcode `\#12\catcode `\^12\catcode `\_12\catcode `\%12\relax}%
\providecommand \@@startlink[1]{}%
\providecommand \@@endlink[0]{}%
\providecommand \url  [0]{\begingroup\@sanitize@url \@url }%
\providecommand \@url [1]{\endgroup\@href {#1}{\urlprefix }}%
\providecommand \urlprefix  [0]{URL }%
\providecommand \Eprint [0]{\href }%
\providecommand \doibase [0]{http://dx.doi.org/}%
\providecommand \selectlanguage [0]{\@gobble}%
\providecommand \bibinfo  [0]{\@secondoftwo}%
\providecommand \bibfield  [0]{\@secondoftwo}%
\providecommand \translation [1]{[#1]}%
\providecommand \BibitemOpen [0]{}%
\providecommand \bibitemStop [0]{}%
\providecommand \bibitemNoStop [0]{.\EOS\space}%
\providecommand \EOS [0]{\spacefactor3000\relax}%
\providecommand \BibitemShut  [1]{\csname bibitem#1\endcsname}%
\let\auto@bib@innerbib\@empty
\bibitem [{\citenamefont {Weinberg}(1995)}]{WeinbergText}%
  \BibitemOpen
  \bibfield  {author} {\bibinfo {author} {\bibfnamefont {S.}~\bibnamefont
  {Weinberg}},\ }\href@noop {} {\emph {\bibinfo {title} {{The Quantum Theory of
  Fields, Vol. I}}}}\ (\bibinfo  {publisher} {Cambridge University Press,
  Cambridge, UK},\ \bibinfo {year} {1995})\BibitemShut {NoStop}%
\end{thebibliography}%

\end{document}


\title{Diffusive Nambu-Goldstone modes in quantum time-crystals: Supplemental material}

\author{Tomoya Hayata}
\affiliation{
Department of Physics, Chuo University, 1-13-27 Kasuga, Bunkyo, Tokyo, 112-8551, Japan 
}
\author{Yoshimasa Hidaka}
\affiliation{
Nishina Center, RIKEN, Wako, Saitama 351-0198, Japan}
\affiliation{
iTHEMS Program, RIKEN, Wako, Saitama 351-0198, Japan
}


\maketitle
\section{Schwinger-Keldysh formalism in an open quantum sytem}
\begin{figure}[htbp] 
   \centering
   \includegraphics[width=0.85\linewidth]{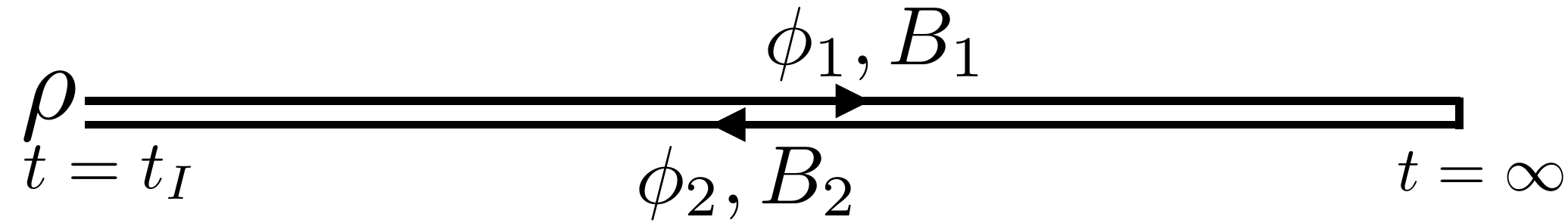} 
   \caption{Closed time contour in the Schwinger-Keldysh formalism.}
   \label{fig:path}
\end{figure}
We here derive the path integral formula for an open quantum system used in the main text.
We consider a quantum system coupled with an environment.
The path integral formula for an open quantum system is obtained by integrating the environment out.
In general, the action of the total system has three parts:
\be
\begin{split}
S_\text{tot}[\vphi,B] = S_\text{sys}[\vphi] + S_\text{env}[B] + S_\text{int}[\vphi,B],
\end{split}
\ee
where $S_\text{sys}[\vphi]$, $S_\text{env}[B]$, and $S_\text{int}[B,\vphi]$ are the actions of  the system,  environment, and interaction, respectively.
$\vphi$ and $B$ are the degrees of freedom of the system and environment.
We assume that the initial density matrix at $t=t_{I}$ is the direct product of those of the system and environment: $\hat\rho= \hat\rho_\text{sys}\otimes \hat\rho_\text{env}$.
We consider the following path integral whose path is shown in Fig.~\ref{fig:path}:
\be
\begin{split}
Z&=  \int \mathcal{D}\vphi_{1}\mathcal{D}\vphi_{2}  \mathcal{D}B_{1}\mathcal{D}B_{2} \rho_\text{sys}(\vphi_{I1},\vphi_{I2})\rho_\text{env}(B_{I1},B_{I2})\\
&\quad\times e^{iS_\text{tot}[\vphi_{1},B_{1}]-iS_\text{tot}[\vphi_{2},B_{2}]},
\label{eq:SK}
\end{split}
\ee
where the indices, $1$ and $2$, represent the label of the forward and backward paths in Fig.~\ref{fig:path},
and we introduced the matrix elements of the density operators $\rho_\text{sys}(\vphi_{I1},\vphi_{I2}):=\langle \vphi_{1}(t_{I})|\hat\rho_\text{sys}|\vphi_{2}(t_{I}) \rangle$
and $\rho_\text{env}(B_{I1},B_{I2}):=\langle B_{1}(t_{I})|\hat\rho_\text{env}|B_{2}(t_{I}) \rangle$.
The direct product of the density operators enables us to formally integrate the environment out.
Then, introducing 
\be
\begin{split}
e^{i\Gamma[\vphi_{1},\vphi_{2}]}&:= \int   \mathcal{D}B_{1}\mathcal{D}B_{2}\rho_\text{env}(B_{I1},B_{I2})\\
&\quad\times e^{iS_\text{env}[B_{1}]-iS_\text{env}[B_{2}]+iS_\text{int}[\vphi_{1},B_{1}]-iS_\text{int}[\vphi_{2},B_{1}]},
\end{split}
\ee
 we can express the path integral of the system as
\be
\begin{split}
Z=\int \mathcal{D}\vphi_{1}\mathcal{D}\vphi_{2}\rho_\text{sys}(\vphi_{I1},\vphi_{I2}) e^{iS_\text{eff}[\vphi_{1},\vphi_{2}]},
\end{split}
\ee
where we define
\be
\begin{split}
S_\text{eff}[\vphi_{1},\vphi_{2}]:=S_\text{sys}[\vphi_{1}]-S_\text{sys}[\vphi_{2}]+\Gamma[\vphi_{1},\vphi_{2}].
\end{split}
\ee
An observable ${O}(\vphi_{1},\vphi_{2})$ of the system is evaluated as
\begin{equation}
\begin{split}
&\langle {O}(\vphi_{1},\vphi_{2}) \rangle\\
&\quad:=
\frac{1}{Z}\int \mathcal{D}\vphi_{1}\mathcal{D}\vphi_{2}\rho_\text{sys}(\vphi_{I1},\vphi_{I2}) e^{iS_\text{eff}[\vphi_{1},\vphi_{2}]} {O}(\vphi_{1},\vphi_{2}).
\end{split}
\end{equation}
When we take $t_{I}\to -\infty$, $\rho_\text{sys}(\vphi_{I1},\vphi_{I2})$ and ${O}(\vphi_{1},\vphi_{2})$ will be uncorrelated.
Then, we can write the path integral as the form in Eq.~(6) in the main text (by generalizing to complex fields).

The $R/A$ basis used in the main text is defined as $\vphi_{R}:=(\vphi_{1}+\vphi_{2})/2$ and $\vphi_{A}:=(\vphi_{1}-\vphi_{2})$.
For example, the action of the $\lambda\vphi^{4}$ theory has the form:
\be
\begin{split}
S_\text{sys}[\vphi] = \int d^{4}x\Bigl(\frac{1}{2}(\partial_{t}\vphi)^{2}-\frac{1}{2}(\nabla\vphi)^{2} - \frac{\lambda}{2} \vphi^{4}\Bigr).
\end{split}
\ee
In the $R/A$ basis, $S_\text{sys}[\vphi_{1}]-S_\text{sys}[\vphi_{2}]$ is expressed as
\be
\begin{split}
&S_\text{sys}[\vphi_{1}]-S_\text{sys}[\vphi_{2}] \\
&=  \int d^{4}x\Bigl(
\vphi_{A}(-\partial_{t}^{2}\vphi_{R}+\nabla^{2}\vphi_{R}-2\lambda\vphi_{R}^{3})-\frac{\lambda}{2}\vphi_{R}\vphi_{A}^{3}
\Bigr)\\
&\quad+ \text{boundary term}.
\end{split}
\ee
This corresponds to the action~(13) in the main text with $\gamma=A=0$.
The functional form of $\Gamma[\vphi_{1},\vphi_{2}]$ depends on the details of the environment. In our van der Pol model in the main text, we chose
\be
\begin{split}
\Gamma[\vphi_{1},\vphi_{2}] = \int d^{4}x\Bigl[
\vphi_{A}\gamma(1-\kappa \vphi^{2}_{R})\partial_{t}\vphi_{R} + i A\vphi_{A}^{2}
\Bigr]  .
\end{split}
\ee

\section{Noether current}
Let us consider an action $S$ as a functional of complex scalar fields $\vPhi_{R}$ and $\vPhi_{A}$.
Suppose that the action $S$ is invariant under an infinitesimal transformation
 $\vPhi_{R}\to \vPhi_{R}+\epsilon_{a}\delta_{A}^{a}\vPhi_{R}$, and   $\vPhi_{A}\to \vPhi_{A}+\epsilon_{a}\delta_{A}^{a}\vPhi_{A}$,
 where $\epsilon_{a}$ is an infinitesimal small constant.
For example,  the  time translation and $U(1)$ transformations are written as 
\be
\begin{split}
\delta_{A}^{T}\vPhi_{R}&:= \partial_{t}\vPhi_{R},\qquad
\delta_{A}^{T}\vPhi_{A}:= \partial_{t}\vPhi_{A},
\end{split}
\ee
and
\be
\begin{split}
\delta_{A}^{Q}\vPhi_{R}&:=i \vPhi_{R},\qquad
\delta_{A}^{Q}\vPhi_{A}:=i \vPhi_{A},
\end{split}
\ee
respectively.
For spacetime-dependent parameters $\epsilon_{a}(x)$, the action behaves as
\be
\begin{split}
S\to S- \int d^{4}x \bigl(\partial_{\mu}\epsilon_{a}(x) \bigr)j_{A}^{a\mu}(x),
\end{split}
\ee
because the last term must vanish for constant $\epsilon_{a}$.
Here, $j_{A}^{a\mu}$ is the so-called Noether current~\cite{WeinbergText}.
For the $U(1)$ symmetry of the action~(17) in the main text, the explicit form of $j_{A}^{Q0}(x)$ is given as
\be
\begin{split}
j_{A}^{Q0} &= i(\vPhi_A^*(\partial_t+i\mu)\vPhi_R-(\partial_t-i\mu)\vPhi_A^* \vPhi_R)\\
&\quad+ i( \vPhi_R^*(\partial_t+i\mu)\vPhi_A -(\partial_t-i\mu)\vPhi_R^*\vPhi_A) \\
&\quad+i\gamma(1-\kappa|\vPhi_R|^2)(\vPhi_R^*\vPhi_A-\vPhi_{A}^{*}\vPhi_R)\\
&\quad+i\lambda ( \vPhi_R \partial_{t}\vPhi_R^{*} -\partial_{t}\vPhi_R \vPhi_R^{*} )( \vPhi_R^*\vPhi_A+ \vPhi_A^*\vPhi_R).
\end{split}
\ee
Similarly, the  charge density of the time translation reads
\be
\begin{split}
j_A^{T0} &= 
- \partial_{t}\vPhi_A^*\partial_{t}\vPhi_R-\partial_{t}\vPhi_R^* \partial_{t}\vPhi_A+ \vPhi_A^*\nabla^{2}\vPhi_R
 +\vPhi_R^*\nabla^{2}\vPhi_A\\
&\quad +(\mu^{2}\!-\!m^{2}\!+\!\lambda|\partial_{t}\vPhi_R|^{2})(\vPhi_A^* \vPhi_R+\vPhi_R^*\vPhi_A)
\!+\! i A|\vPhi_A|^{2}.
\end{split}
\ee
We can introduce the $R$ transformation such that
$\delta_{R}^{a}\vPhi_{A}:= \delta_{A}^{a}\vPhi_{R}$, and $\delta_{R}^{a}\vPhi_{R}:= \delta_{A}^{a}\vPhi_{A}/4$.
Using this, we define the generalized Watanabe-Brauner matrix as $\rho^{ab}=-\langle\delta_{R}^{a}j_{A}^{b0}\rangle$. 
For the superfluid kink time-crystal~(17) in the main text, $\rho^{ab}$ is written explicitly as 
\begin{align}
\langle\delta_{R}^{T}j_{A}^{0T}\rangle&=
(\mu^{2}-m^{2}+\lambda|\dot\vPhi_0|^{2})\partial_{t}|{\vPhi}_0|^{2} 
-\partial_{t}(\dot{\vPhi}_0^*\dot\vPhi_0), \label{eq:delTT}\\
\langle\delta_{R}^{Q}j_{A}^{T0}\rangle&=0,\\
\langle\delta_{R}^{T}j_{A}^{Q0}\rangle&=
i\partial_{t}({\vPhi}_0^*\dot\vPhi_0-\dot\vPhi_0^*\vPhi_0+2i\mu{\vPhi}_0^*\vPhi_0) \notag\\
&\quad-i\gamma\left(1-\kappa|\vPhi_0|^2\right)(\dot\vPhi_{0}^{*}\vPhi_0-\vPhi_{0}^{*}\dot{\vPhi}_{0})\notag\\
&\quad+\lambda i( \vPhi_0 \dot\vPhi_0^{*} -\dot\vPhi_0 \vPhi_0^{*} )\partial_{t}| \vPhi_0|^{2},\\
\langle\delta_{R}^{Q}j_{A}^{Q0}\rangle&=
-2\gamma\left(1-\kappa|\vPhi_0|^2\right)|\vPhi_0|^{2}.  \label{eq:delQQ}
\end{align}
Similarly, for the van der Pol oscillator~(13) in the main text, the explicit form of the charge density of the time translation and  $\langle\delta^{T}_{R}j_{A}^{T0}\rangle$ are, respectively, written as
\begin{align}
j_{A}^{T0} &=-\partial_{t}\vphi_A\partial_{t}\vphi_R
+ \vphi_A \left(\nabla^2-2\lambda\vphi_R^2\right)\vphi_R\notag\\
&\quad-\frac{\lambda}{2}\vphi_{A}^{3}\vphi_{R}
+i A(\vphi_A)^2,\\
 \langle\delta^{T}_{R}j_{A}^{T0}\rangle& =-\partial_{t}\Bigl(\frac{1}{2}\dot{\vphi}_0^{2}
+ \lambda\frac{1}{2}\vphi_0^4\Bigr).
\end{align}

\section{Perturbative solutions}
\label{sec:preliminary}
We here describe the method to compute the dispersion relations of the NG modes in quantum time-crystals.
We would like to solve a second rank linear ordinary differential equation:
\be
\begin{split}
D(t,\partial_{t},\bm{p}) \phi_{\bm{p}}(t) = 0,
\label{eq:ODE}
\end{split}
\ee
where $\bm{p}$ is a parameter. In our situation, $\bm{p}$ is momentum.
Suppose that $D(t,\partial_{t},\bm{p})$ has the time-translational symmetry: $D(t+T,\partial_{t},\bm{p})=D(t,\partial_{t},\bm{p})$, with the period $T$.
Because the rank is two, there are two solutions $\phi_{1,\bm{p}}(t)$ and $\phi_{2,\bm{p}}(t)$. The translational symmetry implies that $\phi_{1,\bm{p}}(t+T)$ and $\phi_{2,\bm{p}}(t+T)$ can be expressed as a linear combination of $\phi_{1,\bm{p}}(t) $ and $\phi_{2,\bm{p}}(t) $:
\be
\begin{split}
\begin{pmatrix}
\phi_{1,\bm{p}}(t+T) \\
\phi_{2,\bm{p}}(t+T) 
\end{pmatrix}
=A
\begin{pmatrix}
\phi_{1,\bm{p}}(t)\\
\phi_{2,\bm{p}}(t)
\end{pmatrix},
\end{split}
\ee
with
\be
\begin{split}
A=\begin{pmatrix}
 c_{11}  &c_{12}\\
c_{21}  &c_{22}
\end{pmatrix}.
\end{split}
\ee
We note that $A$ generally depends on the parameter $p$.
If the eigenvalues of $A$ are nondegenerate, we can diagonalize $A$, with  bases $\tilde{\phi}_{a,p}(t)$, i.e.,
$\tilde{\phi}_{a,\bm{p}}(t+T) = \lambda_{a}\tilde{\phi}_{a,\bm{p}}(t)$, where $a=1,2$. 
In the following we drop the index $a$.
Obviously, $\lambda$ describes a large time behavior. We can parametrize $\lambda$ as $\lambda= e^{-i \omega T}$.
In general, $\omega$ is complex, whose imaginary part is negative if the system is stable.
We can write $\tilde{\phi}_{p}(t)$ as
\be
\begin{split}
\tilde{\phi}_{\bm{p}}(t) = e^{-i \omega t} u_{\bm{p}}(t),
\end{split}
\ee
where $u_{\bm{p}}(t+T)=u_{\bm{p}}(t)$. 
The equation~\eqref{eq:ODE} reads 
\be
\begin{split}
D(t,\partial_{t}-i \omega,\bm{p}) u_{\bm{p}}(t) = 0.
\end{split}
\ee
Let us try to perturbatively solve this equation. 
For this purpose, we decompose $D(t,\partial_{t}-i \omega,\bm{p})$ into 
 $H_{0} = D(t,\partial_{t},\bm{0})$ and $V:=D(t,\partial_{t}-i \omega,\bm{p}) -D(t,\partial_{t},0,\bm{0}) $.
 Then, the equation is written as 
\be
\begin{split}
(H_{0}+V(\omega,\bm{p}))u_{\bm{p}}(t)=0.
\label{eq:c}
\end{split}
\ee
Let $\Psi_{n}(t)$ be eigenfunctions with the eigenvalues $E_{n}$ of $H_{0}$ ($E_0=0$) with the periodic boundary condition $\Psi_{n}(t+T)=\Psi_{n}(t)$,
i.e., 
\be
\begin{split}
H_{0}\Psi_{n}(t) = E_{n}\Psi_{n}(t).
\end{split}
\ee
Since $H_{0}$ is generally not hermite, we also need to introduce $\eta_{n}$ such that
\be
\begin{split}
H^{\dag}_{0}{\eta}_{n}(t) = {E}_{n}{\eta}_{n}(t),
\end{split}
\ee
with
\be
\begin{split}
\int_{0}^{T} dt \eta^{\dag}_{n}(t)\Psi_{m}(t) = \delta_{n,m}.
\end{split}
\ee
Using this basis, we expand $u_{\omega,\bm{p}}(t)$ as
\be
\begin{split}
u_{\omega,\bm{p}}(t) = \sum_{n=0}^{\infty} c_{n}(\omega,\bm{p})\Psi_{n}(t).
\end{split}
\ee
The condition of $c_{n}$ for the zero mode is  at $\omega =0$, $\bm{p}=\bm{0}$, $ c_{0}(0,\bm{0})=1$ and $ c_{n}(0,\bm{0})=0$.
We have
\be
\begin{split}
(H_{0} + V(\omega,\bm{p}))u_{\omega,\bm{p}}(t) &=
\\
\sum_{n=0}^{\infty} c_{n}(\omega,\bm{p})(E_{n}+ V(\omega,\bm{p}) )\Psi_{n}(t)&=0 .
\end{split}
\label{eq:c2}
\ee
Multiplying $\eta_{m}^{\dag}(t)$ and integrating Eq.~\eqref{eq:c2} with respect to $t$, we find
\be
\begin{split}
c_{m}(\omega,\bm{p})E_{m} +  \sum_{n}V_{mn}(\omega,\bm{p}) c_{n}(\omega,\bm{p}) =0,
\end{split}
\ee
where
\be
\begin{split}
V_{mn}(\omega,\bm{p}):=\int_{0}^{T} dt \eta_{m}^{\dag}(t) V(\omega,\bm{p}) \Psi_{n}(t).
\end{split}
\ee
At the leading expansion of $\omega$ and $\bm{p}$,  the solution of Eq.~\eqref{eq:c} is obtained from $V_{00}(\omega,\bm{p})=0$; it gives the relation between $\omega$ and $\bm p$.

\section{Van der Pol oscillator}
The dispersion relation of the NG mode is obtained from the solution of $[ (\rho -\dot{g})\partial_{t}  +g(-\partial_{t}^{2}+\nabla^{2}) ]\chi=:D_{R}^{-1}\chi = 0$,
with $g=\langle\delta_{A}^{T}\vphi_{R}\rangle^{2}= \dot\vphi_{0}^{2}$, and $\rho = -\langle\delta_{R}^{T}j^{T0}\rangle=
\partial_{t}( \dot\vphi_0^{2}/2
+ \lambda\vphi_0^4/2)=\dot{\vphi}_{0}^{2}\gamma(1-\kappa\vphi_{0}^{2})$. Here we used the equation of motion of $\vphi_{0}$ (Eq.~(14) in the main text).
We introduce $G_{R}^{-1}$ such that  $D_{R}^{-1}=\dot{\vphi}_{0}G_{R}^{-1}\dot{\vphi}_{0}$, where 
\be
\begin{split}
& G^{-1}_{R}(t,\partial_t,\nabla)=
\\
&-\partial^2_t+\nabla^2+\gamma\left(1-\kappa\vphi_0^2\right)\partial_t-2\gamma\kappa\vphi_0\dot{\vphi}_0-6\lambda\vphi_0^2 .\;\;
\end{split}
\ee
We can consider the equation $G_{R}^{-1}\delta\phi=0$ instead of  $D_{R}^{-1}\chi= 0$. Obviously, both carry the same information.
For the technical reason of numerics, it is easier to find the solution of $G_{R}^{-1}\delta\phi=0$, rather than that of $D_{R}^{-1}\chi= 0$.
Therefore, in the following, we solve $G_{R}^{-1}\delta\phi=0$ on the basis of the method described in Sec.~\ref{sec:preliminary}.
We  take $D(t,\partial_t,\bm p)=G_{R}^{-1}(t,\partial_t,\bm p)$, $H_0:=G_{R}^{-1}(t,\partial_t,\bm{0})$, and $V(\omega,\bm{p}):=D(t,\partial_t-i\omega,\bm p)-D(t,\partial_t,\bm{0})$, where
\be
H_0=-\partial^2_t+\gamma\left(1-\kappa\vphi_0^2\right)\partial_t-2\gamma\kappa\vphi_0\dot{\vphi}_0-6\lambda\vphi_0^2.
\label{eq:left_van}
\ee
Then, we define $\Psi_0$, and $\eta_0$ as
\begin{align}
H_0\Psi_0&=0  ,\\
H_0^\dagger\eta_0&=0  ,
\label{eq:left_van}
\end{align}
with
\be
H_0^\dagger=-\partial^2_t-\gamma\left(1-\kappa\vphi_0^2\right)\partial_t-6\lambda\vphi_0^2 .
\ee
By construction, we find $\Psi_{0}=\dot{\vphi}_{0}$.
We numerically solve Eq.~\eqref{eq:left_van} under the periodic boundary condition, and compute the left zero-eigenvector $\eta_0$.
Then, by multiplying $\eta_0^{\dag}$ to the left-hand-sides of $V\Psi_0$,
we perturbatively obtain the potential $V_{00}$ as
\be
\begin{split}
V_{00}(\omega,\bm p) &=\int_0^T dt\;\eta_0^{\dag}V(\omega,\bm{p})\dot{\vphi}_{0}\\
&=\int_0^T dt\;  {\psi}_{0}^{\dag} (-i\omega (\rho-\dot{g}) +  g(\omega^{2}-\bm{p}^{2}) )
\end{split}
\ee
where ${\psi}_0^{\dag}= \eta_0^{\dag}/\dot{\vphi}_{0}$.
The normalization condition turns out to be
\begin{equation}
\begin{split}
1= \int_{0}^{T} dt \eta^{\dag}_{0}\Psi_{0} =  \int_{0}^{T} dt \psi^{\dag}_{0}g.
\end{split}
\end{equation}
The on-shell condition is solved as $V_{00}(\omega,\bm p)=0$, which is nothing but Eqs.~(15), and~(16) in the main text.

\section{Superfluid kink time-crystal}
The time periodic solution of Eq.~(18) in the main text, is not strictly periodic but rather quasi-periodic: $\vPhi_{0}(t+T)=e^{i\Omega T}\vPhi_{0}(t)$,
with constant $\Omega$ and $T$, and the linearized equation of motion of fluctuations is also quasi-periodic. In order to apply our technique in which the periodicity of the linearized equation
is assumed,  we redefine the fields such that the expectation value of $\vPhi_{0}(t)$ is periodic, i.e.,
$\vPhi_{R}(x)\to  e^{-i\Omega t}\vPhi_{R}(x)$, and 
$\vPhi_{A}(x)\to  e^{-i\Omega t}\vPhi_{A}(x)$. Under the transformations, the time derivative transforms as $\partial_{t}\to \partial_{t}+i\Omega$, and the linearized equation of motion of fluctuations becomes periodic. We work in this basis in the following analysis.

As the same as before, the dispersion relation of the NG mode is obtained from the solution of $[ (\rho^{ab} -\dot{g}^{ab})\partial_{t}  +g^{ab}(-\partial_{t}^{2}+\nabla^{2}) ]\chi_{b}=[D_{R}^{-1}\chi]_{b} = 0$, where $g^{ab}=2\re\langle\delta_{A}^{a}\vPhi_{R} \rangle\langle\delta_{A}^{b}\vPhi_{R}^{*}\rangle$, with $\langle\delta_{A}^{T}\vPhi_{R}\rangle =\dot\vPhi_{0}$ and  $\langle\delta_{A}^{Q}\vPhi_{R}\rangle =i\vPhi_{0}$.
The explicit expression of $\rho^{ab}$ is shown in Eqs.~\eqref{eq:delTT}-\eqref{eq:delQQ}.
Since both of the time-translation and $U(1)$ symmetries are spontaneously broken,  $D_{R}^{-1}$ is a two-by-two matrix.
Because of the same reason as before, we introduce $G_{R}^{-1}(t,\partial_t,\nabla)$ such that 
\be
\begin{split}
D_{R}^{-1} =
U^{T}G_{R}^{-1}(t,\partial_t,\nabla)U,
\end{split}
\ee
where
\be
\begin{split}
U=
\begin{pmatrix}
\re\langle\delta_{A}^{T}\vPhi_{R}\rangle & \re\langle\delta_{A}^{Q}\vPhi_{R}\rangle\\
\im\langle\delta_{A}^{T}\vPhi_{R}\rangle&\im\langle\delta_{A}^{Q}\vPhi_{R}\rangle
\end{pmatrix}
=\begin{pmatrix}
 \dot{\vphi}_{10}  & -\vphi_{20}  \\ 
\dot{\vphi}_{20}  & \vphi_{10}
\end{pmatrix},
\end{split}
\ee
and
\begin{align}
&[G_{R}^{-1}]_{11}=
\notag \\
&-\partial_t^2+\nabla^2
+\gamma\left(1-\kappa_1\left(\vphi_{10}^2+\vphi_{20}^2\right)\right)\partial_t 
-2\gamma\kappa_1\vphi_{10}\dot{\vphi}_{10}
\notag \\
&+\mu^2-m^2-\lambda\left({\dot{\vphi}_{10}^2+\dot{\vphi}_{20}^2}\right)-2\lambda\vphi_{10}\dot{\vphi}_{10}\partial_t ,
\\
&[G_{R}^{-1}]_{12}=2\mu\partial_t-2\lambda\vphi_{10}\dot{\vphi}_{20}\partial_t-2\gamma\kappa_1\dot{\vphi}_{10}\vphi_{20} ,
\\
&[G_{R}^{-1}]_{21}=-2\mu\partial_t-2\lambda\dot{\vphi}_{10}\vphi_{20}\partial_t-2\gamma\kappa_1\vphi_{10}\dot{\vphi}_{20} ,
\\
&[G_{R}^{-1}]_{22}=
\notag \\
&-\partial_t^2+\nabla^2
+\gamma\left(1-\kappa_1\left(\vphi_{10}^2+\vphi_{20}^2\right)\right)\partial_t -2\gamma\kappa_1\vphi_{20}\dot{\vphi}_{20}
\notag \\
&+\mu^2-m^2-\lambda\left({\dot{\vphi}_{10}^2+\dot{\vphi}_{20}^2}\right)-2\lambda\vphi_{20}\dot{\vphi}_{20}\partial_t .
\end{align}
Here, we parametrized $\vPhi_{0}=\vphi_{10}+i \vphi_{20}$.
To obtain these expressions, we employed the equation of motion of $\vPhi_0$ (Eq.~(18) in the main text).

Now, we take $D(t,\partial_t,\bm p)=M^{-1}G_{R}^{-1}(t,\partial_t,\bm p)M$, $H_0:=D(t,\partial_t,\bm{0})$, and $V(\omega,\bm p):=D(t,\partial_t-i\omega,\bm p)-D(t,\partial_t,\bm{0})$, where
\be
M=
\begin{pmatrix}
\cos \Omega t & -\sin\Omega t \\
\sin\Omega t & \cos \Omega t
\end{pmatrix} .
\ee
$D(t,\partial_t,\bm p)$ satisfy $D(t+T,\partial_t,\bm p)=D(t,\partial_t,\bm p)$ in the new basis rotated by $M$.
The zero-eigenvalue equations of $H_{0}$ and $H_{0}^{\dag}$ satisfy
\begin{align}
H_{0} \tilde{\Psi}^{(i)} &=H_{0} \begin{pmatrix}
\tilde{\Psi}^{(i)}_{10} \\ \tilde{\Psi}^{(i)}_{20}
\end{pmatrix}=0,\\
H_{0}^{\dag} \tilde{\eta}^{(i)} &=
H_0^\dagger
\begin{pmatrix}
\tilde{\eta}^{(i)}_{10} \\ \tilde{\eta}^{(i)}_{20}
\end{pmatrix}=0,
\label{eq:left0}
\end{align}
with the normalization
\be
\int_0^T dt \tilde{\eta}_0^{\dag(i)}\tilde{\Psi}_{0}^{(j)}=\int_0^T dt \eta_0^{\dag(i)}\Psi_{0}^{(j)}=\delta_{ij},
\ee
where $i, j = T,Q$, $\tilde{\Psi}_{0}^{(i)}=M^{-1}\Psi_{0}^{(i)}$, and $\tilde{\eta}_0^{\dag(i)}=M^{-1}\eta_0^{\dag(i)}$.
By construction, we find $\Psi_{0}^{(T)} =(\dot{\vphi}_{10},\dot{\vphi}_{20})$, and $\Psi_{0}^{(Q)}=(-\vphi_{20},\vphi_{10})$.
We numerically solve Eq.~\eqref{eq:left0} under the periodic boundary condition, and compute the left zero-eigenvectors $\tilde{\eta}^{(T)}$ and $\tilde{\eta}^{(Q)}$.
Then, we can perturbatively obtain $V_{00}(\omega,\bm p)$, which is now a two-by-two matrix:
\be
\begin{split}
[V_{00}(\omega,\bm p)]_{ij} &=\int_0^T dt\;\tilde{\eta}^{(i)\dag}V(\omega,\bm{p})\tilde{\Psi}^{(j)}\\
&=\int_0^T dt\;\eta^{(i)\dag}MV(\omega,\bm{p})M^{-1}\Psi^{(j)}\\
&=\begin{pmatrix}
\omega^2+ia\omega-\bm p^2 & -ic\omega \\
ib\omega & \omega^2+id\omega-\bm p^2
\end{pmatrix},
\end{split}
\ee
where
\be
\begin{split}
a&=-\gamma+2\int_0^T dt\left(\eta_{10}^{(T)} \ddot{\vphi}_{10}+\eta_{20}^{(T)} \ddot{\vphi}_{20}\right)
\\
&\quad+\gamma\kappa\int_0^T dt\left(\eta_{10}^{(T)}\dot{\vphi}_{10}+\eta_{20}^{(T)}\dot{\vphi}_{20}\right)\left(\vphi_{10}^2+\vphi_{20}^2\right)
\\
&\quad+2\lambda\int_0^T dt\left(\eta_{10}^{(T)}\vphi_{10}+\eta_{20}^{(T)}\vphi_{20}\right)\left(\dot{\vphi}_{10}^2+\dot{\vphi}_{20}^2\right)
\\
&\quad-2\mu\int_0^T dt\left(\eta_{10}^{(T)}\dot{\vphi}_{20}-\eta_{20}^{(T)}\dot{\vphi}_{10}\right) ,
\end{split} 
\ee
\be
\begin{split}
b&=2\int_0^T dt\left(\eta_{10}^{(Q)} \ddot{\vphi}_{10}+\eta_{20}^{(Q)} \ddot{\vphi}_{20}\right)
\\
&\quad+\gamma\kappa\int_0^T dt\left(\eta_{10}^{(Q)}\dot{\vphi}_{10}+\eta_{20}^{(Q)}\dot{\vphi}_{20}\right)\left(\vphi_{10}^2+\vphi_{20}^2\right)
\\
&\quad+2\lambda\int_0^T dt\left(\eta_{10}^{(Q)}\vphi_{10}+\eta_{20}^{(Q)}\vphi_{20}\right)\left(\dot{\vphi}_{10}^2+\dot{\vphi}_{20}^2\right)
\\
&\quad-2\mu\int_0^T dt\left(\eta_{10}^{(Q)}\dot{\vphi}_{20}-\eta_{20}^{(Q)}\dot{\vphi}_{10}\right) ,
\end{split} 
\ee
\be
\begin{split}
c&=-2\int_0^T dt\left(\eta_{10}^{(T)} (-\dot{\vphi}_{20})+\eta_{20}^{(T)} \dot{\vphi}_{10}\right)
\\
&\quad-\gamma\kappa\int_0^T dt\left(\eta_{10}^{(T)}(-\vphi_{20})+\eta_{20}^{(T)}\vphi_{10}\right)\left(\vphi_{10}^2+\vphi_{20}^2\right)
\\
&\quad-2\lambda\int_0^T dt\left(\eta_{10}^{(T)}\vphi_{10}+\eta_{20}^{(T)}\vphi_{20}\right)\left(-\vphi_{20}\dot{\vphi}_{10}+\vphi_{10}\dot{\vphi}_{20}\right)
\\
&\quad+2\mu\int_0^T dt\left(\eta_{10}^{(T)}\vphi_{10}-\eta_{20}^{(T)}(-\vphi_{20})\right) ,
\end{split} 
\ee
\be
\begin{split}
d&=-\gamma+2\int_0^T dt\left(\eta_{10}^{(Q)} (-\dot{\vphi}_{20})+\eta_{20}^{(Q)} \dot{\vphi}_{10}\right)
\\
&\quad+\gamma\kappa\int_0^T dt\left(\eta_{10}^{(Q)}(-\vphi_{20})+\eta_{20}^{(Q)}\vphi_{10}\right)\left(\vphi_{10}^2+\vphi_{20}^2\right)
\\
&\quad+2\lambda\int_0^T dt\left(\eta_{10}^{(Q)}\vphi_{10}+\eta_{20}^{(Q)}\vphi_{20}\right)\left(-\vphi_{20}\dot{\vphi}_{10}+\vphi_{10}\dot{\vphi}_{20}\right)
\\
&\quad-2\mu\int_0^T dt\left(\eta_{10}^{(Q)}\vphi_{10}-\eta_{20}^{(Q)}(-\vphi_{20})\right) .
\end{split} 
\ee
The on-shell condition $\det V_{00}(\omega,\bm p)=0$ can be solved, 
and the solution at small $\bm p$ takes the form given in Eq.~(20) in the main text.

\bibliography{./ng.bib}